\documentclass[11pt]{article}
\usepackage[dvipdfmx]{color}
\usepackage{graphicx}
\usepackage{xcolor}
  \usepackage{amsthm,amsfonts}
  \usepackage{amsmath}
\usepackage{mathabx}
\usepackage{slashed}
\usepackage{ulem}
\usepackage{cite}

\def\zzg{${\mathbb Z}_2\times{\mathbb Z}_2$-graded }

\newcommand{\bea}   {\begin{eqnarray}}
\newcommand{\eea}   {\end{eqnarray}}

\begin{document}
\renewcommand{\thefootnote}{\fnsymbol{footnote}}

\thispagestyle{empty}
\title{New aspects of the  ${\mathbb Z}_2\times {\mathbb Z}_2$-graded $1D$ superspace:\\
induced strings and $2D$ relativistic models}
\author{N. Aizawa\thanks{{E-mail: {\it aizawa@omu.ac.jp}}}, \quad  R. Ito\thanks{{E-mail: {\it sd27009y@st.omu.ac.jp}}}, \quad Z. Kuznetsova\thanks{{E-mail: {\it zhanna.kuznetsova@ufabc.edu.br}}} \quad and\quad
F.
Toppan\thanks{{E-mail: {\it toppan@cbpf.br}}}
\\
\\
}
\maketitle

\centerline{$^{\ast \dag}$ {\it Department of Physics, Graduate School of Science,}}{\centerline{\it  Osaka Metropolitan University,}}{\centerline{\it Nakamozu Campus, Sakai, Osaka 599-8531, Japan.}}
\centerline{$^{\ddag}$ {\it UFABC, Av. dos Estados 5001, Bangu,}}\centerline{\it { cep 09210-580, Santo Andr\'e (SP), Brazil.}}
{\centerline{$^{\S}$ 
{\it CBPF, Rua Dr. Xavier Sigaud 150, Urca,}}\centerline{\it{
cep 22290-180, Rio de Janeiro (RJ), Brazil.}}
~\\
\maketitle
\begin{abstract}
~\\
A novel feature of the ${\mathbb Z}_2\times {\mathbb Z}_2$-graded supersymmetry which finds no counterpart in ordinary supersymmetry is the presence of $11$-graded exotic bosons (implied by the existence of two classes of parafermions). Their interpretation, both physical and mathematical, presents a challenge. The role of the ``exotic bosonic coordinate" was not considered by previous works on the one-dimensional  \zzg superspace (which was restricted to produce point-particle models).
By treating this coordinate at par with the other graded superspace coordinates new consequences are obtained. \\
The graded superspace calculus of
the \zzg worldline super-Poincar\'e algebra induces two-dimensional \zzg relativistic models; they are invariant under a new \zzg  $2D$ super-Poincar\'e algebra which differs from the previous  two \zzg $2D$ versions of super-Poincar\'e introduced in the literature. In this new superalgebra the second translation generator and the Lorentz boost are
$11$-graded. Furthermore, if the exotic coordinate is compactified on a circle ${\bf S}^1$,  a \zzg closed string with periodic boundary conditions is derived. \\
The analysis of the irreducibility conditions of the $2D$ supermultiplet implies that a larger $(\beta$-deformed, where $\beta\geq 0$ is a real parameter) class of point-particle models than the ones discussed so far in the literature (recovered at $\beta=0$)  is obtained. While the spectrum of the $\beta=0$ point-particle models is degenerate (due to its relation with an  ${\cal N}=2$ supersymmetry), this is no longer the case for the $\beta> 0$ models.
\end{abstract}
\vfill
\rightline{CBPF-NF-004/22}
\newpage

\section{Introduction}

This paper presents a careful investigation of the properties of the graded superspace associated with the \zzg worldline super-Poincar\'e algebra. The main focus is the role of the ``exotic bosonic coordinate" which was not considered by previous works on \zzg superspace since it was constrained to produce point-particle models. The $11$-graded exotic bosonic coordinate (the superalgebra generators and superspace coordinates belong to the $00,~ 10,~ 01$ and $11$ graded sectors) is the novel feature
which finds no counterpart in the ordinary (i.e., ${\mathbb Z}_2$-graded) superspace. Therefore its interpretation, 
both mathematical and physical, is a non-trivial challenge. In this paper we treated the exotic coordinate at par with the other graded superspace coordinates (they are given by an ordinary boson which plays the role of time  and two parafermions). We anticipate the main consequences. The calculus implied by the extra, exotic, bosonic coordinate induces two-dimensional \zzg relativistic models. They are invariant under a new two-dimensional \zzg super-Poincar\'e algebra; unlike the two previous \zzg two-dimensional super-Poincar\'e algebras introduced in \cite{bruz2n} and in \cite{tol}, for this new superalgebra the second translation generator and the Lorentz boost are $11$-graded. Furthermore, if the exotic coordinate is compactified on a circle ${\bf S}^1$,  a \zzg closed string with periodic boundary conditions is derived. \par
Before introducing in more detail the further results obtained in the paper, we briefly sketch the state-of-the-art of the investigations on \zzg Lie superalgebras and their physical applications. \par
\zzg Lie algebras and superalgebras were introduced in \cite{{riwy1},{riwy2},{sch1}} as the simplest ${\mathbb Z}_2^n$-graded extensions of ordinary Lie (super)algebras. The \zzg Lie algebras involve parabosons,
see e.g. \cite{kuto}. The \zzg Lie superalgebras on the other hand contain as subalgebras the ${\mathbb Z}_2$-graded Lie superalgebras presented in  \cite{kac}; they involve (para)fermions and for this reason they can be naturally applied to generalizations of supersymmetric theories. Early works discussing physical applications of \zzg Lie superalgebras are \cite{{luri},{vas},{jyw}}.\par
In recent years \zzg Lie superalgebras found a renewed attention due to progresses in different directions. It became clear that superalgebras of this class are symmetries of well-known physical models such as, see \cite{{aktt1},{aktt2}}, the non-relativistic L\'evy-Leblond spinors. Systematic investigations of \zzg invariant theories, both classical \cite{akt1,akt2,brusigma,bruSG} and quantum \cite{brdu,AAD},  appeared;  conformally invariant models and more general ${\mathbb Z}_2^n$-graded invariant theories, for $n\geq 2$, have started been 
investigated \cite{DoiAi1,AAd2}.  It was further shown, see \cite{{top1},{top2}}, that the paraparticles implied by \zzg theories are theoretically detectable and lead to physical consequences (for previous works on \zzg parastatistics see \cite{{YJ,tol1},{stvj1}}).\par
On the mathematical side various studies of algebraic and geometric aspects of ${\mathbb  Z}_2^n$-graded structures have been investigated since their introduction. 
	Here we mention only very recent works of algebraic studies which discuss structures and representations  \cite{AmaAi,que,NeliJoris,LuTan} (further references on algebraic aspects are found in \cite{AmaAi}). 
	The differential geometry on $\mathbb{Z}_2^n$-graded manifolds, which has a close kinship with the \zzg superspace formulation, is also a field of extensive study; for details one can see the concise reviews \cite{pon,PonSch}.  

Concerning the \zzg superspace formulation, this topic has been investigated in \cite{brusigma,aido1,aido2}.\par
We highlight some further results, besides the ones already mentioned at the beginning, of this work. We point out in particular the extension of the \cite{aido1} analysis of the irreducibility of the supermultiplets.  As a consequence  we derive $\beta$-deformed (in terms of a real parameter $\beta$) worldline $D$-module representations
of the  \zzg $1D$ super-Poincar\'e algebra. The representations investigated in the literature  in \cite{aido1} (which lead to invariant models possessing a supersymmetric spectrum) are recovered at the special, undeformed, $\beta=0$ point.
We also introduce a convenient matrix representation of the graded supercoordinates; it
allows to reconstruct the \zzg calculus from matrices (which encode the ${\mathbb Z}_2\times{\mathbb Z}_2$ grading) coupled to the Berezin's calculus \cite{ber}. This representation simplifies the construction of invariant actions.
More comments on this work and the future perspectives are given in the Conclusions.\par
The scheme of the paper is the following. In Section {\bf 2} the basic features of \zzg Lie superalgebras are recalled.
The graded superspace is introduced in Section {\bf 3}. Section {\bf 4} presents the graded superfields.
The recovering of irreducible supermultiplets is discussed in Section {\bf 5}.  The three types of $D$-module representations of the \zzg worldline super-Poincar\'e algebra are given in Section {\bf 6}. Section {\bf 7} outlines the construction of the graded invariant actions. The matrix representation of the superspace is given in Section {\bf 8}. The induced two-dimensional \zzg super-Poincar\'e algebra is derived in Section {\bf 9}.
Two-dimensional relativistic actions, invariant under the  \zzg super-Poincar\'e algebra, are obtained in Section {\bf 10}. In Section {\bf 11} it is shown that the compactification on a ${\bf S}^1$ circle of the exotic bosonic coordinate leads to a \zzg closed string with periodic boundary conditions. The last Section presents a summary of the results, outlining the main open questions and further lines of investigation. In the Appendix the irreducible, $\beta$-deformed worldline generators are derived.

\section{${\mathbb Z}_2\times {\mathbb Z}_2$-graded superalgebras}

In this Section we recall the basic features of ${\mathbb Z}_2\times {\mathbb Z}_2$-graded superalgebras and
introduce the relevant cases (the one-dimensional ${\mathbb Z}_2\times {\mathbb Z}_2$-graded super-Poincar\'e algebra and the graded abelian superalgebra) that are discussed in the following.\par
A ${\mathbb Z}_2\times {\mathbb Z}_2$-graded Lie superalgebra ${\cal G}$  (for our purposes the graded superalgebra is defined over the ${\mathbb R}$, ${\mathbb C}$ fields of real or
complex numbers) is an extension of an ordinary superalgebra which is decomposed according to
\bea
{\cal G} &=& {\cal G}_{00}\oplus {\cal G}_{01}\oplus {\cal G}_{10}\oplus {\cal G}_{11};
\eea 
the grading $\vec{\alpha}= deg (a)$ of a generator $a\in{\cal G}$ is given by the pair $\vec{\alpha}^T=(\alpha_1,\alpha_2)$, with $\alpha_1,\alpha_2\in \{0,1\}$. 
The bracket $(\cdot,\cdot ):{\cal G}\times{\cal G}\rightarrow {\cal G}$ is introduced on homogeneous generators $a,b$ of grading $\vec{\alpha},\vec{\beta}$ through the position
\bea\label{rbracket}
(a,b)&=& ab -(-1)^{{\vec\alpha}\cdot{\vec \beta}}ba,\qquad {\textrm{where}}\quad 
{\vec\alpha}\cdot{\vec\beta} = \alpha_1\beta_1+\alpha_2\beta_2\quad mod ~2.
\eea
Depending on the $(-1)^{{\vec\alpha}\cdot{\vec\beta}}$ sign, the above bracket coincides with a commutator or an anticommutator; when specified, throughout the paper commutators (anticommutators) are denoted as
 $[\cdot,\cdot]$ (and, respectively, $\{\cdot,\cdot\}$).\par
The grading of the $(a,b)$ generator is given by
\bea\label{mod2sum}
deg((a,b)) &=& {\vec{\alpha}+\vec{\beta}}\quad mod ~ 2.
\eea
For any three elements $a,b,c\in {\cal G}$ of respective grading ${\vec{\alpha}, {\vec\beta}},{\vec{\gamma}}$ the following graded Jacobi identity is satisfied:
\bea
\label{gradedjacobi}
 (-1)^{\vec{\gamma}\cdot\vec{\alpha}}(a,(b,c))+
 (-1)^{\vec{\alpha}\cdot\vec{\beta}}(b,(c,a))+
 (-1)^{\vec{\beta}\cdot\vec{\gamma}}(c,(a,b))&=&0.
\eea
The graded superalgebras under consideration here are: \par
~\\
{{\it i}) {\it{the one-dimensional ${\mathbb Z}_2\times {\mathbb Z}_2$-graded super-Poincar\'e algebra ${\cal P}$}}}
which is
defined, see \cite{brdu,akt1,akt2}, in terms of $4$ generators whose grading assignment is
$H\in {\cal P}_{00},~Q_{10}\in {\cal P}_{10},~ Q_{01}\in {\cal P}_{01}, ~Z\in {\cal P}_{11} $ and whose nonvanishing (anti)commutators are
\bea\label{z2z2poin}
\{Q_{10},Q_{10}\} = \{Q_{01},Q_{01}\} = 2H, &\quad& [Q_{10}, Q_{01}]= iZ
\eea
(as shown in Section {\bf 9}, the addition of an extra $11$-graded ``Lorentz-boost" generator
consistently defines a two-dimensional  \zzg super-Poincar\'e algebra ${\cal P}_{d=2}$);
~\\
{\it ii})  {\it{the graded abelian algebra ${\cal A}$ of the superspace coordinates}}}, whose $4$ generators are denoted as 
$t\in {\cal A}_{00},~\theta_{10}\in {\cal A}_{10},~ \theta_{01}\in {\cal A}_{01}, ~z\in {\cal A}_{11} $. All its (anti)commutators defined by ({\ref{rbracket}}) are vanishing; this implies, in particular, that
$\theta_{10}, \theta_{01} $ are nilpotent ($\theta_{10}^2=\theta_{01}^2=0$).
\par
~\\
{\it  iii}) {\it{{the larger graded superalgebra ${\cal S}$, spanned by }}}
\bea\label{larger}
&H,t\in {\cal S}_{00},~~~Q_{10},\theta_{10}\in {\cal S}_{10},~~~ Q_{01},\theta_{01}\in {\cal S}_{01}, ~~~Z,z\in {\cal S}_{11}. &
\eea
The superalgebras ${\cal P}, {\cal A}$ are recovered as subalgebras (${\cal P}, {\cal A}\subset {\cal S}$); in ${\cal S}$ the (anti)commutators $(a,p)$ for $a\in {\cal A}$, $p\in {\cal P}$  are defined to be all vanishing. The further superalgebra extension ${\cal S}'$ accommodates the derivatives $\partial_t, \partial_{10}, \partial_{01},\partial_z$ of the (respective) superspace coordinates. The grading of these derivatives, whose action is defined in the next Section, is given by
\bea\label{gradedder}
&\partial_t\in{\cal S}'_{00},\quad  \partial_{10}\in{\cal S}'_{10},\quad \partial_{01}\in{\cal S}'_{01},\quad \partial_z\in{\cal S}'_{11}.&
\eea
\par
~\\
The operation of star conjugation $\ast$, which allows to define the hermitian operators,  is defined for graded generators $a,b$ and $\lambda\in {\mathbb C}$ to satisfy
\bea\label{star}
&(ab)^\ast = b^\ast a^\ast, \quad (a^\ast)^\ast =a,\quad (\lambda a)^\ast =\lambda^\ast a^\ast,&
\eea
where $\lambda^\ast $ denotes the complex conjugation of $\lambda$.\par
Throughout the paper we assume the generators of ${\cal S}$ to be hermitian, so that
\bea\label{herm}
&H^\ast=H, ~~Q_{10}^\ast=Q_{10},~~Q_{01}^\ast=Q_{01},~~ Z^\ast = Z, ~~t^\ast = t, ~~ \theta_{10}^\ast=\theta_{10},~~\theta_{01}^\ast=\theta_{01},~~z^\ast=z.&
\eea
The construction here presented is a particular case of the differential calculus \cite{ckp} on ${\mathbb Z}_2^n$-supermanifolds.

\section{The graded superspace}

The graded superspace of the ${\mathbb Z}_2\times {\mathbb Z}_2$-graded one-dimensional
super-Poincar\'e algebra (\ref{z2z2poin}) is recovered by introducing a group element $g\in G$ by setting
\bea\label{groupelement}
g &=& \exp (itH-\theta_{10}Q_{10}-\theta_{01}Q_{01}+izZ)
\eea
in terms of the graded super-Poincar\'e generators and superspace coordinates entering (\ref{larger}).  It follows, by taking into account the hermiticity conditions (\ref{herm}), that $g$ is unitary
\bea
g^{-1} &=& g^{\ast}.
\eea
A transformation $g\mapsto g'\in G$ is derived from the left action
\bea\label{gleft}
\qquad g' &=& g_\epsilon \cdot g,\qquad {\textrm{with}} \qquad 
g_\epsilon = \exp (i\epsilon_{00}H-\epsilon_{10}Q_{10}-\epsilon_{01}Q_{01}+i\epsilon_{11}Z),
\eea
for the infinitesimal $\epsilon_{ij}$ ($\epsilon_{ij}^\ast=\epsilon_{ij}$) graded coordinates.\\
Let us collectively denote with $X\in\{t, \theta_{10},\theta_{01}, z\}$ the graded superspace coordinates, so that $g\equiv g(X)$ and $g'\equiv g(X')$ for $X'=X + \delta X$. The transformations of the graded superspace coordinates are given by
\bea\label{transf1}
\delta t = \epsilon_{00}+ i\epsilon_{10}\theta_{10}+i\epsilon_{01}\theta_{01}, &\quad&~~ \delta z = \epsilon_{11}+\frac{1}{2}(\epsilon_{10}\theta_{01}-\epsilon_{01}\theta_{10}),\nonumber\\
\delta \theta_{10} = \epsilon_{10},\qquad\qquad\qquad\quad  ~~~&\quad& \delta\theta_{01} =\epsilon_{01}.
\eea
The above transformations are obtained from the Baker-Campbell-Hausdorff formula. They can also be recovered from the left action of differential  operators of the graded
superspace coordinates  (\ref{gradedder}). The nonvanishing left action of the derivatives
on the graded superspace coordinates are given by
\bea\label{gradedderiv}
&\partial_{10}\cdot \theta_{10}=1,\qquad
\partial_{01}\cdot \theta_{01}= 1,\qquad \partial_z \cdot z^k = k z^{k-1},
\eea
while $\partial_t\equiv \frac{\partial}{\partial t}$ is the ordinary partial derivative of the ordinary real coordinate $t$.\par

By equating (\ref{transf1}) with  the transformations expressed by
\bea\label{deltax}
\delta X&= &(-i\epsilon_{00}{\widehat H} +\epsilon_{10}{\widehat Q}_{10}+\epsilon_{01}{\widehat Q}_{01}-i\epsilon_{11}{\widehat Z}) X \qquad {\textrm{for}} \quad X\in \{t, \theta_{10},\theta_{01}, z\},
\eea
one gets
\bea\label{operators}
{\widehat H} =i \partial_t,\qquad\qquad\qquad \qquad&\quad&~~{\widehat Z} = i\partial_z,\nonumber\\
{\widehat Q}_{10} = \partial_{10}+i\theta_{10}\partial_t+\frac{1}{2}\theta_{01}\partial_z,&\quad& {\widehat Q}_{01} =\partial_{01} +i \theta_{01}\partial_t-\frac{1}{2}\theta_{10}\partial_z.
\eea
The above set of operators gives a differential representation of the graded super-Poincar\'e algebra (\ref{z2z2poin}) where, in particular, the nonvanishing (anti)commutators are
\bea\label{poirep}
\{{\widehat Q}_{10},{\widehat Q}_{10}\} = \{{\widehat Q}_{01},{\widehat Q}_{01}\} = 2{\widehat H}, &\quad& [{\widehat Q}_{10}, {\widehat Q}_{01}]= i {\widehat Z}.
\eea
By construction, the above operators are hermitian:
\bea
&{\widehat H}^\dagger ={\widehat H},\qquad {\widehat Q}_{10}^\dagger ={\widehat Q}_{10},\qquad {\widehat Q}_{01}^\dagger ={\widehat Q}_{01},\qquad {\widehat Z}^\dagger ={\widehat Z}.&
\eea
The (para)fermionic covariant derivatives ${\widehat D}_{10}, {\widehat D}_{01}$ are obtained from the right actions. They are given by
\bea\label{covder}
{\widehat D}_{10} = \partial_{10}-i\theta_{10}\partial_t-\frac{1}{2}\theta_{01}\partial_z,&\quad& {\widehat D}_{01} =\partial_{01} -i \theta_{01}\partial_t+\frac{1}{2}\theta_{10}\partial_z.
\eea
The covariant derivatives satisfy the (anti)commutators \bea\label{covardev1}
\{{\widehat D}_{10},{\widehat D}_{10}\} = \{{\widehat D}_{01},{\widehat D}_{01}\} = -2{\widehat H}, &\quad& [{\widehat D}_{10}, {\widehat D}_{01}]= -i {\widehat Z}
\eea
and 
\bea\label{covardev2}
&\{{\widehat D}_{10},{\widehat Q}_{10}\} = \{{\widehat D}_{01},{\widehat Q}_{01}\} =
[{\widehat D}_{10},{\widehat Q}_{01}] = [{\widehat D}_{01},{\widehat Q}_{10}] =0.&
\eea

\section{${\mathbb Z}_2\times {\mathbb Z}_2$-graded superfields}

A real, ${\mathbb Z}_2\times {\mathbb Z}_2$-graded superfield $\Phi(t,\theta_{10},\theta_{01},z)$ admits a decomposition, by taking into account the nilpotency of $\theta_{10}, \theta_{01}$ and the special properties of $z$, in terms of $8$ component fields denoted as $\varphi_{00}, ~{\widetilde \varphi}_{00},~\varphi_{11}, ~{\widetilde \varphi}_{11},~\psi_{10}, ~{\widetilde \psi}_{10},~\psi_{01}, ~{\widetilde \psi}_{01}$ (the suffix indicates their respective gradings). The component fields are functions of the real coordinate $t$ and of the real parameter $x$ introduced as
\bea
\qquad\qquad\qquad x= z^2 &&{\textrm{($x\geq 0$~~due~to~the~hermiticity~of~$z$).}}
\eea

The superfield decomposition is
\bea\label{decomposition}
\Phi(t,\theta_{10},\theta_{01},z)&=& 1\cdot \bigl( \varphi_{00}(t,x)+z{\widetilde \varphi}_{11}(t,x) \bigr) +\theta_{10}\cdot\bigl( i\psi_{10}(t,x)+z{\widetilde \psi}_{01}(t,x)\bigr) +\nonumber\\&&\theta_{01}\cdot\bigl( i \psi_{01}(t,x)+z{\widetilde \psi}_{10}(t,x)\bigr) +\theta_{10}\theta_{01}\cdot \bigl(\varphi_{11}(t,x)+z{\widetilde \varphi}_{00}(t,x)\bigr).
\eea
By taking into account the reality properties (\ref{herm}), the reality condition
\bea\label{sfieldreality}
\Phi(t,\theta_{10},\theta_{01},z)^\ast&=&
\Phi(t,\theta_{10},\theta_{01},z)
\eea
implies that the $8$ component fields entering (\ref{decomposition}) are all real.\par
By normalizing the scaling dimension to be $[H]=1$, it follows from (\ref{z2z2poin}) and (\ref{groupelement}) that the scaling dimensions of the 
one-dimensional ${\mathbb Z}_2\times {\mathbb Z}_2$-graded super-Poincar\'e generators and their superspace coordinates are given by
\bea
[H]=[Z]=1, \qquad [Q_{10}]=[Q_{01}]=\frac{1}{2}, &~~& [t]=[z]=-1,\qquad [\theta_{10}]=[\theta_{01}]=-\frac{1}{2}.
\eea
It turns out, once assumed 
$[\Phi(t,\theta_{10},\theta_{01},z)]=s$, that the scaling dimensions of the component fields are given by
\bea\label{scalings}
s&:& \varphi_{00},\nonumber\\
s+\frac{1}{2}&:&\psi_{10}, ~\psi_{01},\nonumber\\
s+1
&:& \varphi_{11},~ {\widetilde \varphi}_{11},\nonumber\\
s+\frac{3}{2}&:& {\widetilde \psi}_{10},~{\widetilde \psi}_{01},\nonumber\\
s+2&:&{\widetilde \varphi}_{00},\eea
while the scaling dimensions of their coordinates are
\bea
[t]=-1, && [x]=-2.
\eea

Let us set, as before,  $X\equiv t, \theta_{10},\theta_{01}, z$; the scalar property of the superfield $\Phi(X)$,
\bea
\Phi'(X') &=& \Phi(X), \qquad \quad {\textrm{for}}\quad X'=X+\delta X,
\eea
where $\delta X$ is given in (\ref{deltax}), implies
\bea
&\delta\Phi(X)=\Phi(X')-\Phi(X)= (i\epsilon_{00}{\widehat H} -\epsilon_{10}{\widehat Q}_{10}-\epsilon_{01}{\widehat Q}_{01}+i\epsilon_{11}{\widehat Z}) \Phi(X).&
\eea
The transformations of the component fields can be read from the action of the ${\widehat H},~{\widehat Q}_{10},~{\widehat Q}_{01},~{\widehat Z}$ operators presented in (\ref{operators}). When applied to the $8$-dimensional
vector $v$ given by 
\bea\label{gradedvector}
v^T &=&(\varphi_{00}, {\widetilde\varphi}_{00},\varphi_{11}, {\widetilde\varphi}_{11}, \psi_{10}, {\widetilde\psi}_{10},\psi_{01}, {\widetilde\psi}_{01})
\eea
the (\ref{operators}) operators produce a $8\times 8$ differential matrix representation in $t,x$ of the  graded superalgebra (\ref{poirep}).  The corresponding matrix operators, denoted with a regular instead of the italic font
used in (\ref{operators}), are
\bea\label{diffreptx}
{\widehat{\textrm{H}}}~&=&i\partial_t\cdot {\mathbb I}_8,\nonumber\\
{\widehat{\textrm{Q}}}_{10}&=&{\footnotesize\left(\begin{array}{cccccccc} 
0&0&0&0&i&0&0&0\\
0&0&0&0&-i\partial_x&i\partial_t&0&0\\
0&0&0&0&0&0&-\partial_t&-\frac{1}{2}-x\partial_x\\
0&0&0&0&0&0&0&1\\
\partial_t&0&0&0&0&0&0&0\\
\partial_x&1&0&0&0&0&0&0\\
0&0&-i&-\frac{i}{2}-ix\partial_x&0&0&0&0\\
0&0&0&i\partial_t&0&0&0&0
\end{array}\right)},\nonumber\\
~~{\widehat {\textrm{Q}}}_{01}&=&
{\footnotesize\left(\begin{array}{cccccccc} 
0&0&0&0&0&0&i&0\\
0&0&0&0&0&0&i\partial_x&i\partial_t\\
0&0&0&0&-\partial_t&\frac{1}{2}+x\partial_x&0&0\\
0&0&0&0&0&1&0&0\\
0&0&-i&\frac{i}{2}+ix\partial_x&0&0&0&0\\
0&0&0&i\partial_t&0&0&0&0\\
\partial_t&0&0&0&0&0&0&0\\
-\partial_x&1&0&0&0&0&0&0
\end{array}\right)},\nonumber\\
{\widehat {\textrm{Z}}}~&=&{\footnotesize{\left(\begin{array}{cccccccc} 
0&0&0&i+2ix\partial_x&0&0&0&0\\
0&0&2i\partial_x&0&0&0&0&0\\
0&i+2ix\partial_x&0&0&0&0&0&0\\
2i\partial_x&0&0&0&0&0&0&0\\
0&0&0&0&0&0&0&-1-2x\partial_x\\
0&0&0&0&0&0&2\partial_x&0\\
0&0&0&0&0&-1-2x\partial_x&0&0\\
0&0&0&0&2\partial_x&0&0&0
\end{array}\right)}}
\eea
(we denote, here and in the following, the $n\times n$ Identity matrix with the symbol ${\mathbb I}_n$).\par
The diagonal operator ${\widehat {\textrm Z}}^2$, 
\bea
{\widehat{\textrm Z}}^2&=& -(2\partial_x+4x\partial_x^2)\cdot{\mathbb I}_8 -4\partial_x\cdot(E_{22}+E_{44}+E_{66}+E_{88}),
\eea 
is a Casimir operator of the (\ref{z2z2poin}) superalgebra (here and in the following $E_{ij}$ denotes the matrix with entries $1$ at the crossing of the $i$-th row with the $j$-th column and $0$ otherwise). \par
The representations are labeled by the  eigenvalues of 
$ {\widehat{\textrm Z}}^2$. Since these eigenvalues are non-negative, they can be expressed as $\lambda^2$, in terms of a real parameter $\lambda$ which, without loss of generality, can be assumed to be $\lambda\geq 0$. The scaling dimension of $\lambda$ is
\bea
[\lambda] &=&1.
\eea
In physical applications the parameter $t$ is identified with the time, while ${\widehat {\textrm H}}$ is the Hamiltonian operator. The physical interpretation of $x$ as en extra dimension induced  by the exotic
 bosonic coordinate will be discussed in the following. 
The component fields entering  (\ref{gradedvector}) are defined in  $(1+1)$-dimensions; the graded superfield can be associated with the symbol below which
specifies the numbers of component fields, see formula (\ref{scalings}), of respective scaling dimension $s,s+{\footnotesize\frac{1}{2}},s+1,s+{\footnotesize\frac{3}{2}}, s+2$:
\bea
&(1;2;2;2;1).&
\eea
Analogous symbols have been employed to describe representations of one-dimensional supermechanics \cite{pato,krt}  and of one-dimensional ${\mathbb Z}_2\times {\mathbb Z}_2$-graded mechanics \cite{akt1}.\par
{The $(1+1)$-dimensional extended ${\mathbb Z}_2\times {\mathbb Z}_2$-graded superspace generalizes the previous constructions of  one-dimensional
${\mathbb Z}_2\times {\mathbb Z}_2$-graded superspace presented in \cite{pon, aido1}.
The results of these works are recovered by suitably constraining the extended supermultiplets.
These constraints, which are based on the notion of irreducible supermultiplet, are only applicable when point-particle mechanics can be derived. In the following we present a more detailed analysis of the admissible point-particle constraints. \par The Hamiltonian
${\widehat{\textrm H}}$ is a Casimir operator. Its energy eigenvalue $E\geq 0$, with scaling dimension $
[E]= 1$, produces together with $\lambda\geq 0$ the pair of values
\bea
&(E,\lambda)&
\eea
which label, see \cite{aido1,aido2}, the irreducible representations. They are given by
\bea\label{energylambda}
~I&:&\quad (E\geq 0, \lambda =0) \quad~~~~ {\textrm{and}}\nonumber\\
  II&:&\quad (E=\alpha\lambda, \lambda>0) .
\eea
As shown in Appendix {\bf A} the parameter $\alpha$ (which in principle is real and non-negative) has to satisfy
the constraint
\bea
\label{alphaonehalf}
\alpha &\geq& \frac{1}{2}.
\eea

\section{Irreducible graded supermultiplets}

The general forms of the $8$ eigenfunctions corresponding to the $
{\widehat{\textrm Z}}^2=\lambda^2$ eigenvalue are
\bea\label{eigenfunctions}
{\textrm{at}}\quad \lambda=0:\qquad\qquad&&\nonumber\\
&& \qquad f_A(t) +f_B(t){\textstyle{\sqrt{x}}}\quad ~~{\textrm{for}}\quad \varphi_{00},\varphi_{11},\psi_{10},\psi_{01},\nonumber\\&&
\qquad f_A(t) +f_B(t){\textstyle{{\frac{1}{\sqrt{x}}}}}\quad ~~ {\textrm{for}}\quad {\widetilde \varphi}_{00},{\widetilde \varphi}_{11},{\widetilde \psi}_{10},{\widetilde \psi}_{01};\nonumber\\
&&\nonumber\\
{\textrm{at}}\quad \lambda\neq0:\qquad\qquad&&\nonumber\\
&&\qquad  f_A(t)\cos(\lambda{\sqrt x}) +f_B(t)\sin(\lambda{\sqrt{x}})\quad ~~~{\textrm{for}}\quad \varphi_{00},\varphi_{11},\psi_{10},\psi_{01},\nonumber\\&&
{\textstyle{\frac{1}{\sqrt{x}}}}\big(f_A(t)\cos(\lambda{\sqrt x}) +f_B(t)\sin(\lambda{\sqrt{x}})\big)\quad ~~ {\textrm{for}}\quad {\widetilde \varphi}_{00},{\widetilde \varphi}_{11},{\widetilde \psi}_{10},{\widetilde \psi}_{01}.\nonumber\\
&&
\eea
In the above formulas $f_A(t),~ f_B(t)$ are generic functions of $t$.\par
~\par
{\bf 1) The $\lambda=0$ case.} \par
~\par
At $\lambda=0$ the restriction
\bea
{\widehat{\textrm{Z}}}&=& 0
\eea
can be consistently imposed. It is easily checked that this condition (which corresponds to the constraint imposed in \cite{brusigma}) implies:\par
~\par
~{\it i}) the ${\widetilde \varphi}_{00},{\widetilde \varphi}_{11},{\widetilde \psi}_{10},{\widetilde \psi}_{01}$ fields are all vanishing and\par
{\it ii}) the ${ \varphi}_{00},{ \varphi}_{11},{ \psi}_{10},{ \psi}_{01}$ fields have no dependence on $x$. \par
~\par

The action of ${\widehat{\textrm{Q}}}_{10}, {\widehat{\textrm{Q}}}_{01}$ on the restricted vector
$v_{r}^T= \big(\varphi_{00}(t), \varphi_{11}(t), \psi_{10}(t), \psi_{01}(t)\big)$ produces the minimal $4\times 4$ matrix differential representation, see \cite{brdu, akt1}, of the $Z=0$ Beckers-Debergh \cite{bede} algebra. The restricted operators
${\widehat{\textrm{Q}}}_{10}^{(r)}, {\widehat{\textrm{Q}}}_{01}^{(r)}$ are
\bea
{\widehat{\textrm{Q}}}_{10}^{(r)}={\footnotesize\left(\begin{array}{cccc} 
0&0&i&0\\
0&0&0&-\partial_t\\
\partial_t&0&0&0\\
0&-i&0&0
\end{array}\right)}, &&{\widehat{\textrm{Q}}}_{01}^{(r)}={\footnotesize\left(\begin{array}{cccc} 
0&0&0&i\\
0&0&-\partial_t&0\\
0&-i&0&0\\
\partial_t&0&0&0
\end{array}\right)}.
\eea
By setting ${\widehat{\textrm{H}}}^{(r)} = i\partial_t\cdot{\mathbb I}_4$ we get
\bea
&\{{\widehat{\textrm{Q}}}_{10}^{(r)},{\widehat{\textrm{Q}}}_{10}^{(r)}\}=\{{\widehat{\textrm{Q}}}_{01}^{(r)},{\widehat{\textrm{Q}}}_{01}^{(r)}\}=2{\widehat{\textrm{H}}}^{(r)}, \qquad [{\widehat{\textrm{H}}}^{(r)},{\widehat{\textrm{Q}}}_{10}^{(r)}]=[{\widehat{\textrm{H}}}^{(r)},{\widehat{\textrm{Q}}}_{01}^{(r)}]=[{\widehat{\textrm{Q}}}_{10}^{(r)},{\widehat{\textrm{Q}}}_{10}^{(r)}]=0.&
\eea

~\par
A few comments are in order:\par
~ {\it I} - the scaling dimension of $\lambda$ is the same as the one of a mass-term. On the other hand a mass term $m$ can be introduced in a model even when $\lambda=0$;
\par
{\it II} - the reality condition (\ref{sfieldreality}) for the superfield $\Phi(X)$ is valid at the {\it classical} level. In the quantum case, in order to impose the eigenvalue equation for the Hamiltonian, i.e.
\bea
&i\partial_t {\mathbf \Psi} ={\widehat{\textrm H}}{\mathbf \Psi} = E{\mathbf \Psi}&
\eea
where $E$ is the energy, the time coordinate $t$ entering the eigenfunctions (\ref{eigenfunctions}) should be Wick-rotated ($t\rightarrow - it$). In the supersymmetric case the distinction bewteen classical versus quantum $D$-module reps is discussed, e.g.,  in \cite{cht}, while the extension to  the ${\mathbb Z}_2\times {\mathbb Z}_2$-graded invariant setting has been presented in \cite{akt2}.\par
~\par
{\bf 2) The  ${\mathbf \lambda}{\neq 0}$ case.}\par
~\par
At $\lambda\neq 0$ a representation satisfying the $Z^2=\lambda^2$ Casimir is recovered by the action of the (\ref{diffreptx}) operators on the $8$-dimensional vectors expressed, by introducing the separation of the $t,x$ coordinates, as
{\small{\bea\label{newvec}
v_\lambda^T &=& \big(g_{00}(t)h(x),{\widetilde g}_{00}(t){\widetilde h}(x),g_{11}(t)h(x),{\widetilde g}_{11}(t){\widetilde h}(x),
g_{10}(t)h(x),{\widetilde g}_{10}(t){\widetilde h}(x),g_{01}(t)h(x),{\widetilde g}_{01}(t){\widetilde h}(x)\big).\nonumber\\
\eea 
}} 
In the above formula $h(x),~{\widetilde h}(x)$  denote the dimensionless functions
\bea
h(x) = e^{i\lambda{\sqrt x}}, && {\widetilde h}(x) = \frac{1}{\lambda {\sqrt x}}e^{i\lambda{\sqrt x}},
\eea 
while $g_{ij}(t), ~{\widetilde g}_{ij}(t)$ are functions of the time coordinate $t$.\par
The identities
\bea
\partial_x h(x) = \frac{i\lambda^2}{2} {\widetilde h}(x), &\quad&(\frac{1}{2}+x\partial_x){\widetilde h}(x) =\frac{i}{2}h(x)
\eea
imply that, restricting the operators on the (\ref{newvec}) vectors, one obtains an $8\times 8 $ differential matrix realization of the (\ref{z2z2poin}) superalgebra which only depends on the time coordinate $t$.

The resulting operators are
\bea\label{difftrep}
{\widehat{\textrm{H}}^{(R)}}~&=&i\partial_t\cdot {\mathbb I}_8,\nonumber\\
{\widehat{\textrm{Q}}}_{10}^{(R)}&=&{\footnotesize\left(\begin{array}{cccccccc} 
0&0&0&0&i&0&0&0\\
0&0&0&0&\frac{\lambda^2}{2}&i\partial_t&0&0\\
0&0&0&0&0&0&-\partial_t&-\frac{i }{2}\\
0&0&0&0&0&0&0&1\\
\partial_t&0&0&0&0&0&0&0\\
\frac{i\lambda^2}{2}&1&0&0&0&0&0&0\\
0&0&-i&\frac{1}{2}&0&0&0&0\\
0&0&0&i\partial_t&0&0&0&0
\end{array}\right)},\nonumber\\
~~{\widehat {\textrm{Q}}}_{01}^{(R)}&=&
{\footnotesize\left(\begin{array}{cccccccc} 
0&0&0&0&0&0&i&0\\
0&0&0&0&0&0&-\frac{\lambda^2}{2}&i\partial_t\\
0&0&0&0&-\partial_t&\frac{i}{2}&0&0\\
0&0&0&0&0&1&0&0\\
0&0&-i&-\frac{1}{2}&0&0&0&0\\
0&0&0&i\partial_t&0&0&0&0\\
\partial_t&0&0&0&0&0&0&0\\
-\frac{i\lambda^2}{2}&1&0&0&0&0&0&0
\end{array}\right)},\nonumber\\
{\widehat {\textrm{Z}}}^{(R)}~&=&{\footnotesize{\left(\begin{array}{cccccccc} 
0&0&0&-1&0&0&0&0\\
0&0&-\lambda^2&0&0&0&0&0\\
0&-1&0&0&0&0&0&0\\
-\lambda^2&0&0&0&0&0&0&0\\
0&0&0&0&0&0&0&-i\\
0&0&0&0&0&0&i\lambda^2&0\\
0&0&0&0&0&-i&0&0\\
0&0&0&0&i\lambda^2&0&0&0
\end{array}\right)}}.
\eea
The $(E,\lambda)$ representation is obtained by imposing the energy eigenvalue, which means to substitute
$i\partial_t\mapsto E$ in the entries of the above matrices. We set, as in formula (\ref{energylambda}),
\bea\label{ealphalambda}
E&=& \alpha \lambda,
\eea
where the real parameter $\alpha$ is nondimensional.\par
The irreducible representations of the $\mathbb{Z}_2 \times \mathbb{Z}_2$-graded super-Poincar\'e algebra ${\cal P}$ were investigated in \cite{aido1} (see \cite{aido2} for the ${\cal N}=2$ extension of the graded super-Poincar\'e algebra).  
The $4$-dimensional, minimal, $(\alpha\lambda,\lambda)$ irreducible representation is recovered by consistently constraining the $g_{ij}(t), {\widetilde g}_{ij}(t)$ fields entering (\ref{newvec}) in terms of 
$4$ time-dependent fields $f_{00}(t), f_{10}(t), f_{01}(t), f_{11}(t)$ of respective scaling dimensions
$s, s+\frac{1}{2}, s+\frac{1}{2}, s+1$; the resulting representation will therefore be symbolically expressed as $(1;2;1)$. \par
By taking into account the dimensionality of the fields, we are looking for a constrained vector $v_{constr}$ given by
\bea
v_{constr}^T &=& ( f_{00}, ~k_0\lambda^2f_{00},~ k_1f_{11}, ~k_2f_{11},~ f_{10},~ k_3\lambda f_{10},~ f_{01},~ k_4\lambda f_{01});
\eea
the identifications are $g_{00}(t)=f_{00}(t), ~{\widetilde g}_{00}(t) = k_0\lambda^2 f_{00}(t)$ and so on.
The dimensionless constants $k_0, k_1,k_2,k_3$ have to be selected in order to guarantee the compatibility of the transformations obtained from the left actions of (\ref{difftrep}).\
The compatibility conditions imply:
\bea
k_0 &=&\frac{1}{2}\sqrt{4\alpha^2-1},\nonumber\\
k_1 &=&k_2\frac{1-4\alpha^2+i\sqrt{4\alpha^2-1}}{2i-2\sqrt{4\alpha^2-1}},\nonumber\\
k_3 &=&\frac{-1+i\sqrt{4\alpha^2-1}}{2\alpha},\nonumber\\
k_4 &=&\frac{1+i\sqrt{4\alpha^2-1}}{2\alpha}.
\eea
The $k_2\neq 0$ constant is arbitrary. Without loss of generality it can be selected to be
\bea
k_2&=&1.
\eea
The resulting $4$-dimensional irreducible representation is given by
\bea\label{alphalambda4}
{\widehat{\textrm{Q}}}_{10}^{(\alpha)}&=&{\footnotesize\left(\begin{array}{cccc} 
0&0&i&0\\
0&0&0&\frac{1+i\sqrt{4\alpha^2-1}}{2\alpha}\lambda\\
-i\alpha\lambda&0&0&0\\
0&\frac{2\alpha^2}{1+i\sqrt{4\alpha^2-1}}&0&0
\end{array}\right)}, \nonumber\\
{\widehat{\textrm{Q}}}_{01}^{(\alpha)}&=&{\footnotesize\left(\begin{array}{cccc} 
0&0&0&i\\
0&0&\frac{-1+i\sqrt{4\alpha^2-1}}{2\alpha}\lambda&0\\
0&-\frac{1-2\alpha^2+i\sqrt{4\alpha^2-1}}{1+i\sqrt{4\alpha^2-1}}&0&0\\
-i\alpha\lambda&0&0&0
\end{array}\right)},\nonumber\\
{\widehat{\textrm{Z}}}^{(\alpha)}&=&{\footnotesize\left(\begin{array}{cccc} 
0&-1 &0&0\\
-\lambda^2&0&0&0\\
0&0&0&\frac{-i+\sqrt{4\alpha^2-1}}{2\alpha}\lambda\\
0&0&\frac{2\alpha}{-i+\sqrt{4\alpha^2-1}}\lambda&0
\end{array}\right)},
\eea
together with ${\widehat{\textrm{H}}}^{(\alpha)}= \alpha\lambda \cdot{\mathbb I}_4$.\par
~\par

As shown in Appendix {\bf A}, $\alpha$ is constrained to satisfy $\alpha\geq \frac{1}{2}$. The second order Casimir $C_2$, introduced through the position
\bea
C_2 &:=&\big( {\widehat{{\textrm H}}^{(\alpha)}}\big)^2-\frac{1}{4}\big({\widehat{\textrm Z}^{(\alpha)}}\big)^2 \equiv
\big(\alpha^2-\frac{1}{4}\big)\lambda^2,
\eea
is such that it recovers 
\bea
&C_2 = 0 \qquad{\textrm{at the special ``boundary" value}} \quad \alpha=\frac{1}{2}.&
\eea
This special point is of particular importance because it produces, as discussed in Appendix {\bf A},  an (${\cal N}=2$) supersymmetry. The following anticommutators (which are not defined as ${\mathbb Z}_2\times{\mathbb Z}_2$-graded operations, but can nevertheless being computed) are nonvanishing:
\bea
\{ {\widehat{\textrm{Q}}}_{10}^{(R)}, 
{\widehat{\textrm{Q}}}_{01}^{(R)}\}& \neq & 0 \qquad{\textrm{for the $8\times 8$ matrices entering (\ref{difftrep}) and}}
\eea
\bea
\{ {\widehat{\textrm{Q}}}_{10}^{(\alpha >\frac{1}{2})}, 
{\widehat{\textrm{Q}}}_{01}^{(\alpha >\frac{1}{2})}\} &\neq & 0 \qquad{\textrm{for the  $4\times 4$ matrices entering  (\ref{alphalambda4}) with $\alpha>\frac{1}{2}$.}}
\eea
On the other hand, at $\alpha=\frac{1}{2}$, the $4\times4$ matrices $  {\widehat{\textrm{Q}}}_{10}^{(\alpha=\frac{1}{2})},~ {\widehat{\textrm{Q}}}_{01}^{(\alpha=\frac{1}{2})}$ satisfy
\bea\label{susycritical}
\{ {\widehat{\textrm{Q}}}_{10}^{(\alpha=\frac{1}{2})}, 
{\widehat{\textrm{Q}}}_{01}^{(\alpha=\frac{1}{2})}\} &=& 0.
\eea
The spectrum of the  ${\mathbb Z}_2\times{\mathbb Z}_2$-graded invariant models presented in \cite{brdu,akt1,akt2} is supersymmetric since these models were constructed for this special $\alpha=\frac{1}{2}$ value. The $\alpha >\frac{1}{2}$ representations induce non supersymmetric generalizations of ${\mathbb Z}_2\times{\mathbb Z}_2$-graded invariant theories.

\section{Three types of $D$-module representations}
The differential operators (\ref{diffreptx}) induce three types  of $D$-module representations of the one-dimensional
 ${\mathbb Z}_2\times{\mathbb Z}_2$-graded  super-Poincar\'e algebra. They are
\par
~\\
~~{\it i}) the $8$-dimensional representation, symbolically expressed as $(1;2;2;2;1)$, defined for a $1+1$ spacetime described by the time coordinate $t$ and a space coordinate $y$ of scaling dimensions $[t]=[y]=-1$. As discussed in the following, the $1+1$-dimensional spacetime can be assumed to be Minkowski. This $D$-module representation is applied to construct invariant relativistic $2$-dimensional sigma models and string actions;\par
~\\
~ {\it ii}) the class of $4$-dimensional representations, labeled by a real parameter $\beta\geq 0$, which can be symbolically expressed as $(1;2;1)_{\beta}$. They depend on the time coordinate $t$ and find application in the construction of invariant world-line sigma models and point particle quantum mechanics; \par
~\\
{\it iii}) the further class of world-line  $4$-dimensional representations, also labeled by $\beta\geq 0$, which can be symbolically expressed as $(2;2;0)_{\beta}$. With respect to the previous class, these differential operators act on component fields with different scaling dimensions (respectively given by $s,s,s+\frac{1}{2}, s+\frac{1}{2}$).
\par
~
\par
The three types of $D$-module representations are here presented.

\subsection{The $(1;2;2;2;1)$ $D$-module representation for the $1+1$ Minkowski spacetime}

The $(1;2;2;2;1)$ $D$-module representation for the $1+1$ Minkowski spacetime is recovered from the
(\ref{diffreptx}) operators after setting, for $x>0$ and in a convenient normalization,
\bea\label{xy2}
\qquad x ={\footnotesize{ \frac{1}{4}}} y^2, &&  \quad {\textrm{where}}\quad y\in{\mathbb R}.
\eea

The resulting operators, acting on
\bea
{\overline {\mathbf v}}^T &=&{\small{ \big(
\varphi_{00}(t,y), {\widetilde \varphi}_{00}(t,y),
\varphi_{11}(t,y), {\widetilde \varphi}_{11}(t,y),
\psi_{10}(t,y), {\widetilde \psi}_{10}(t,y),
\psi_{01}(t,y), {\widetilde \psi}_{01}(t,y)
\big)
}}
\eea
 and denoted in boldface as ${\overline {\mathbf M}}$, are 
{{\bea\label{reldmod}
{\overline {\mathbf H}}&=& i\partial_t\cdot {\mathbb I}_8,\nonumber\\
{\overline {\mathbf Q}}_{10}&=& {\small{\left(\begin{array}{cccccccc} 
0&0&0&0&i&0&0&0\\
0&0&0&0&-{\frac{2i}{y}}\partial_y&i\partial_t&0&0\\
0&0&0&0&0&0&-\partial_t&-\frac{1}{2}-{\frac{y}{2}}\partial_y\\
0&0&0&0&0&0&0&1\\
\partial_t&0&0&0&0&0&0&0\\
{\frac{2}{y}}\partial_y&1&0&0&0&0&0&0\\
0&0&-i&-\frac{i}{2}-{\frac{iy}{2}}\partial_y&0&0&0&0\\
0&0&0&i\partial_t&0&0&0&0
\end{array}\right)}} ,\nonumber\\
{\overline {\mathbf Q}}_{01}&=&{\small{\left(\begin{array}{cccccccc} 
0&0&0&0&0&0&i&0\\
0&0&0&0&0&0&\frac{2i}{y}\partial_y&i\partial_t\\
0&0&0&0&-\partial_t&\frac{1}{2}+{\frac{y}{2}}\partial_y&0&0\\
0&0&0&0&0&1&0&0\\
0&0&-i&\frac{i}{2}+{\frac{iy}{2}}\partial_y&0&0&0&0\\
0&0&0&i\partial_t&0&0&0&0\\
\partial_t&0&0&0&0&0&0&0\\
-{\frac{2}{y}}\partial_y&1&0&0&0&0&0&0
\end{array}\right) }},\nonumber\\
{\overline {\mathbf Z}}&=&{\small{\left(\begin{array}{cccccccc} 
0&0&0&i+iy\partial_y&0&0&0&0\\
0&0&{\frac{4i}{y}}\partial_y&0&0&0&0&0\\
0&i+iy\partial_y&0&0&0&0&0&0\\
{\frac{4i}{y}}\partial_y&0&0&0&0&0&0&0\\
0&0&0&0&0&0&0&-1-y\partial_y\\
0&0&0&0&0&0&{\frac{4}{y}}\partial_y&0\\
0&0&0&0&0&-1-y\partial_y&0&0\\
0&0&0&0&{\frac{4}{y}}\partial_y&0&0&0
\end{array}\right) .}}
\eea
}}

\subsection{The $(1;2;1)_\beta$ worldline  $D$-module representations}

The $(1;2;1)_\beta$ worldline  $D$-module representations are directly obtained from the $4\times 4$ matrices
(\ref{alphalambda4}) by taking into account that one can restore the dependence of the differential operator $\partial_t$ in the
$i\partial_t\mapsto\alpha\lambda$ mapping. One can therefore set
\bea
&\lambda\equiv \frac{i}{\alpha}\partial_t,\qquad {\textrm{plus the position}}\quad
\alpha=\beta +\frac{1}{2}, \qquad {\textrm{where}} \quad \beta\geq 0.&
\eea
Following the presentation (\ref{alphabeta}) in Appendix {\bf A}, the shifted parameter $\beta $ is introduced in order to get at $\beta=0$ the supersymmetric critical point
from equation (\ref{susycritical}). 
\par
The resulting operators, acting on 
\bea\label{hatmultiplet}
{\widehat {\mathbf v}}^T &=& \big({\widehat f}_{00}(t),{\widehat f}_{11}(t),{\widehat f}_{10}(t),{\widehat f}_{01}(t)\big)
\eea
and denoted in boldface as ${\widehat {\mathbf M}}^\beta$, are 
{{\bea\label{121beta}
{\widehat {\mathbf H}}^\beta&=& i\partial_t\cdot {\mathbb I}_4,\nonumber\\
{\widehat {\mathbf Q}}_{10}^\beta&=&  \left( \begin{array}{cccc} 0&0&i&0\\
0&0&0&\frac{2i-4\sqrt{\beta(\beta+1)}}{(2\beta+1)^2}\partial_t\\
\partial_t&0&0&0\\
0&-\frac{i(2\beta+1)^2}{-2i+4\sqrt{\beta(\beta+1)}}&0&0\\
\end{array}\right),\nonumber\\
{\widehat {\mathbf Q}}_{01}^\beta&=& \left( \begin{array}{cccc} 0&0&0&i\\
0&0&-\frac{2(i+2\sqrt{\beta(\beta+1)})}{(2\beta+1)^2}\partial_t&0\\
0&\frac{i-4i\beta-4i\beta^2-4\sqrt{\beta(\beta+1)}}{-2i+4\sqrt{\beta(\beta+1)}}&0&0\\
\partial_t&0&0&0\\
\end{array}\right) ,\nonumber\\
{\widehat {\mathbf Z}}^\beta&=& \left( \begin{array}{cccc} 0&-1&0&0\\
\frac{4}{(2\beta+1)^2}\partial_t^2&0&0&0\\
0&0&0&\frac{2+4i\sqrt{\beta(\beta+1)}}{(2\beta+1)^2}\partial_t\\
0&0&\frac{2i}{-i+2\sqrt{\beta(\beta+1)}}\partial_t&0\\
\end{array}\right).
\eea}}

\subsection{The $(2;2;0)_\beta$ worldline  $D$-module representations}

The $(2;2;0)_\beta$ worldline  $D$-module representations are obtained from the $(1;2;1)_\beta$ representations by applying a dressing transformation. The $(2;2;0)_\beta$ differential matrices, denoted in boldface as ${\widecheck {\mathbf M}}^\beta$, can be expressed from the corresponding $(1;2;1)_\beta$ operators 
${\widehat {\mathbf M}}^\beta$ as
\bea
{\widehat {\mathbf M}}^\beta&\mapsto & {\widecheck {\mathbf M}}^\beta = {\mathbf D}^{-1} {\widehat {\mathbf M}}^\beta{\mathbf D}, \qquad {\textrm{for}} \quad {\mathbf D}= diag(1, i\partial_t, 1,1).
\eea
The dressing transformation was introduced in \cite{pato}, see also \cite{krt}, for the one-dimensional supermechanics
and in \cite{akt1} for the ${\mathbb Z}_2\times {\mathbb Z}_2$-graded case at the special $\beta=0$ point.\par
The ${\widecheck {\mathbf M}}^\beta$ operators acting on 
\bea
{\widecheck {\mathbf v}}^T &=& \big({\widecheck f}_{00}(t),{\widecheck f}_{11}(t),{\widecheck f}_{10}(t),{\widecheck f}_{01}(t)\big)
\eea
are
{{\bea\label{checkmultiplet}
{\widecheck {\mathbf H}}^\beta&=& i\partial_t\cdot {\mathbb I}_4,\nonumber\\
{\widecheck {\mathbf Q}}_{10}^\beta&=&\left( \begin{array}{cccc} 0&0&i&0\\
0&0&0&-i\frac{2i-4\sqrt{\beta(\beta+1)}}{(2\beta+1)^2}\\
\partial_t&0&0&0\\
0&-i\frac{i(2\beta+1)^2}{-2i+4\sqrt{\beta(\beta+1)}}\partial_t&0&0\\
\end{array}\right) ,\nonumber\\
{\widecheck {\mathbf Q}}_{01}^\beta&=&
 \left( \begin{array}{cccc} 0&0&0&i\\
0&0&i\frac{2(i+2\sqrt{\beta(\beta+1)})}{(2\beta+1)^2}&0\\
0&i\frac{i-4i\beta-4i\beta^2-4\sqrt{\beta(\beta+1)}}{-2i+4\sqrt{\beta(\beta+1)}}\partial_t&0&0\\
\partial_t&0&0&0\\
\end{array}\right) ,\nonumber\\
{\widecheck {\mathbf Z}}^\beta&=& 
\left( \begin{array}{cccc} 0&-i\partial_t&0&0\\
-i\frac{4}{(2\beta+1)^2}\partial_t&0&0&0\\
0&0&0&\frac{2+4i\sqrt{\beta(\beta+1)}}{(2\beta+1)^2}\partial_t\\
0&0&\frac{2i}{-i+2\sqrt{\beta(\beta+1)}}\partial_t&0\\
\end{array}\right).
\eea
}}
~\par
~\par
{\it Comment}: for all three types of $D$-module representations the respective covariant derivatives, obtained from (\ref{covder}) and satisfying (\ref{covardev1},\ref{covardev2}), can be constructed.

\section{On the construction of ${\mathbb Z}_2\times {\mathbb Z}_2$-graded invariant actions}

Before presenting in the next Section the calculus for the graded superspace of the one-dimensional ${\mathbb Z}_2\times {\mathbb Z}_2$-graded super-Poincar\'e algebra, it is worth pointing out
some features of the construction of invariant classical action terms.\par
In formula (\ref{diffreptx}) we can replace $x\geq 0$ with
\bea
&x={\overline y}^r, \qquad {\overline y}=x^{\frac{1}{r}} \qquad {\textrm{so that}}\quad [x]=-2, \quad [{\overline y}]= -\frac{2}{r}.&
\eea
For $r=2$ the coordinate proportional to ${\overline y}$, as shown in (\ref{xy2}),  has the same scaling dimension of the time $t$ ($[t]=[{\overline y}]=-1$)
and can be used in a relativistic construction.   \par
By replacing the $\partial_x$ and $\frac{1}{2} + x \partial_x$ entries in formulas (\ref{diffreptx}) with
\bea
\partial_x &\mapsto & \frac{1}{r} {\overline y}^{1-r}\partial_{\overline y},\nonumber\\
\frac{1}{2} + x \partial_x &\mapsto& \frac{1}{2} +\frac{1}{r}{\overline y}\partial_{\overline y},
\eea
one can check under which condition the integrand
\bea
&a\varphi_{11}+b{\widetilde\varphi}_{11}&
\eea
transforms as a total derivative under the action of ${\widehat {\textrm Q}}_{10}^{(R)} $,  ${\widehat {\textrm Q}}_{01}^{(R)} $, so that
\bea
&\int \int d{\overline y}dt~ \big( a \varphi_{11} (t,{\overline y})+ b{\widetilde \varphi}_{11}(t,{\overline y})\big)&
\eea
 is invariant. We get, up to partial integration terms,
\bea
\delta_{Q_{10}}\big( a \varphi_{11}(t,{\overline y})+ b{\widetilde \varphi}_{11}(t,{\overline y)}\big) &\approx &  -a {\dot \psi}_{01} -\frac{a}{r} ({\overline y}{\widetilde\psi}_{01})' + \frac{(2-r)a+2rb}{2r}{\widetilde\psi}_{01},\nonumber\\
\delta_{Q_{01}}\big( a \varphi_{11}(t,{\overline y})+ b{\widetilde \varphi}_{11}(t,{\overline y})\big) &\approx &  -a {\dot \psi}_{10} -\frac{a}{r} ({\overline y}{\widetilde\psi}_{10})' + \frac{(r-2)a+2rb}{2r}{\widetilde\psi}_{10},
\eea
where the dot and the prime respectively denote the partial derivatives $\partial_t$, $\partial_{\overline y}$.
A total derivative in the right hand side requires the vanishing of both last terms presented in the above equations. The solution
\bea
 b=0, && r=2,
\eea
implies that the invariant contribution to the action can only be constructed for the homogeneous spacetime $[t]=[{{\overline y}}]=-1$ which gives the $D$-module representation (\ref{reldmod}).  We denote as $y$, see (\ref{xy2}), the corresponding normalized space coordinate. The constraint $ b=0$ corresponds to the so-called integrability condition introduced in \cite{pon} and discussed in \cite{aido1}.
\par

If we assume $y$ to be compactified on a circle $y\in [0,2\pi[$, with $\varphi_{11} (t,y)$ even function of $y$
and ${\widetilde \varphi}_{11} (t,y)$ odd function of $y$,
we can
set
\bea\label{compactintegral}
\frac{1}{2\pi}\oint dy\int dt \varphi_{11}(t,y) &=& \sum_{n=0}^\infty \varphi^n_{11} (t) \cos (ny),\qquad{\textrm{so that}}
\nonumber\\
\frac{1}{2\pi}\oint dy\int dt \varphi_{11}(t,y) &=& \int dt \varphi^0_{11}(t)
 \eea
reproduces the integration on the ${\mathbb Z}_2\times{\mathbb  Z}_2$-graded superspace introduced in \cite{pon}. We are able to go further.
\par
By assuming $y\in {\mathbb R}$ and the $y$-dependent component functions to be in ${\cal L}^2({\mathbb R})$,
the invariant term
\bea\label{inv12221}
&\int \int_{-\infty}^{+ \infty} dy ~ dt~ \big( \varphi_{11}(t,y)\big)&
\eea
generalizes the \cite{pon} integration.\par
~\par
For the $(1;2;1)_\beta$ worldline supermultiplet (\ref{hatmultiplet}) the invariant term can be expressed
as
\bea\label{inv121}
&\int dt {\widehat f}_{11}(t).&
\eea
Both (\ref{inv12221}) and (\ref{inv121}) are $11$-graded. \par
The invariant terms of the
$(2;2;0)_\beta$ worldline supermultiplet (\ref{checkmultiplet}) are
\bea
\int dt {\widecheck f}_{10}(t)\quad &{\textrm{and}}&\quad 
\int dt {\widecheck f}_{01}(t).
\eea
They are $10$- and $01$-graded.

\section{Matrix representations of the graded superspace}

The calculus (derivations, integrations) on the ${\mathbb Z}_2\times{\mathbb Z}_2$-graded superspace
is easily described in terms of a matrix representation of the graded superspace coordinates (\ref{larger}), realized by two real parameters $t,y$ and two real Grassmann numbers $\theta,\xi$. It follows, in particular, that the Berezin
calculus is one of the ingredients entering the \zzg calculus.\par
The ${\mathbb Z}_2\times{\mathbb Z}_2$ grading can be accommodated, see \cite{akt1,akt2}, in $4\times 4$ matrices  whose nonvanishing entries of the $[ij]$-sectors are given by 
{\footnotesize{\bea\label{fromz2toz2z2}
 &M_{00}\equiv \left(\begin{array}{cccc}\ast&0&0&0\\ 0&\ast&0&0\\0&0&\ast&0\\ 0&0&0&\ast \end{array}\right),~
 M_{11}\equiv \left(\begin{array}{cccc}0&\ast&0&0\\ \ast&0&0&0\\0&0&0&\ast\\ 0&0&\ast&0 \end{array}\right),~
 M_{10}\equiv \left(\begin{array}{cccc}0&0&\ast&0\\ 0&0&0&\ast\\\ast&0&0&0\\ 0&\ast&0&0 \end{array}\right),~
M_{01}\equiv  \left(\begin{array}{cccc}0&0&0&\ast\\ 0&0&\ast&0\\0&\ast&0&0\\ \ast&0&0&0 \end{array}\right).&\nonumber\\
&&
\eea
}}
One can introduce, as an example of a \zzg algebra of matrices, the (complexified) quaternions, see \cite{kuto}; they are  expressed by ${\mathbb I}_4$ and $ M_i$ for $i=1,2,3$. In terms of the Pauli matrices $\sigma_i$,{\footnotesize{
\bea
&\sigma_1=\left(\begin{array}{cc} 0&1\\1&0\end{array}\right), \qquad \sigma_2=\left(\begin{array}{cc} 0&-i\\i&0\end{array}\right),\qquad \sigma_3=\left(\begin{array}{cc} 1&0\\0&-1\end{array}\right),&
\eea}}
the $M_i$'s are defined through the positions
\bea
&M_1={\mathbb I}_2\otimes\sigma_1,\qquad M_2= \sigma_1\otimes\sigma_2,\qquad M_3=\sigma_1\otimes\sigma_3. &
\eea
They satisfy the relations 
\bea
&M_iM_j =\delta_{ij}{\mathbb I}_4 +i\varepsilon_{ijk} M_k \qquad (M_i^\dagger=M_i),&
\eea
where $\varepsilon_{ijk}$ is the totally antisymmetric tensor with normalization $\varepsilon_{123}=1$. The complexification of the ordinary quaternions imply that the matrices $M_i$ are hermitian.\par
The vanishing (anti)commutators of the graded superspace coordinates $t,z,\theta_{10},\theta_{01}$ are reproduced from the matrix representation
\bea\label{matrixcoor}
&t\mapsto {\overline t} = t\cdot{\mathbb I}_4, \quad \theta_{10}\mapsto {\overline\theta}_{10} =\theta\cdot M_1,\quad
\theta_{01}\mapsto {\overline\theta}_{01} =\xi\cdot M_2,\quad z\mapsto {\overline z} =\frac{1}{2}y\cdot M_3&
\eea
(as mentioned before, $t,y$ are real coordinates while $\theta,\xi$ are real Grassmann numbers).\par
The matrix representation of the graded derivatives (\ref{gradedderiv}) is expressed by the positions
\bea
&\partial_{\overline t} = {\mathbb I}_4\cdot\partial_t, \quad \partial_{\overline z} = 2M_3\cdot\partial_y,\quad
\partial_{\overline{\theta}_{10}} = M_1\cdot\partial_\theta,\quad \partial_{\overline{\theta}_{01}} = M_2\cdot\partial_\xi.&
\eea
It follows that the matrix representation of the \zzg super-Poincar\'e generators (\ref{operators}) and the covariant derivatives ({\ref{covder}) are respectively given by
\bea\label{superspacehatop}
&{\widehat H} = {\widetilde H}\cdot {\mathbb I}_4, \quad  {\widehat Q}_{10} = {\widetilde Q}_1\cdot M_1,\quad
 {\widehat Q}_{01} = {\widetilde Q}_2\cdot M_2, \quad {\widehat Z} = {\widetilde Z}\cdot M_3 &
\eea
and
\bea
\label{superspacecovder2}
& {\widehat D}_{10} = {\widetilde D}_1\cdot M_1,\qquad 
 {\widehat D}_{01} = {\widetilde D}_2\cdot M_2,&
\eea
where
\bea\label{tildeops}
{\widetilde H} = i\partial_t,\qquad\qquad\quad~~&~~&~{\widetilde Z}= 2i\partial_y,\nonumber\\
{\widetilde Q}_1= \partial_\theta+i\theta\partial_t+i\xi\partial_y,&~~&{\widetilde Q}_2= \partial_\xi+i\xi\partial_t+i\theta\partial_y,\nonumber\\
{\widetilde D}_1= \partial_\theta-i\theta\partial_t-i\xi\partial_y,&~~&{\widetilde D}_2= \partial_\xi-i\xi\partial_t-i\theta\partial_y.
\eea
In this way the \zzg operators (\ref{superspacehatop},\ref{superspacecovder2}) are reconstructed in terms of the matrices $M_i$ (which provide the 
${\mathbb Z}_2\times{\mathbb Z}_2$ grading) and the differential operators (\ref{tildeops}). 
The latter operators  close a ${\mathbb Z}_2$-graded superalgebra ${\widetilde {\cal G}}={\widetilde {\cal G}}_0\oplus {\widetilde{\cal G}}_1$ whose even (odd) elements are respectively
\bea
&{\widetilde H},{\widetilde Z}\in{\widetilde {\cal G}}_0, \qquad  {\widetilde Q}_1,  {\widetilde Q}_2, {\widetilde D}_1, {\widetilde D}_2\in {\widetilde {\cal G}}_1.
\eea
The only nonvanishing (anti)commutators of ${\widetilde{\cal G}}$ are
\bea\label{z2alg}
\{{\widetilde Q}_1,{\widetilde Q}_1\}= 
\{{\widetilde Q}_2,{\widetilde Q}_2\}=2{\widetilde H},~~&~~& 
\{{\widetilde Q}_1,{\widetilde Q}_2\}={\widetilde Z},\nonumber\\
\{{\widetilde D}_1,{\widetilde D}_1\}= 
\{{\widetilde D}_2,{\widetilde D}_2\}=-2{\widetilde H},&~~& 
\{{\widetilde D}_1,{\widetilde D}_2\}=-{\widetilde Z}.
\eea
The introduction of the matrices $M_i$ allows to make the passage from (\ref{z2alg}) to the \zzg algebra (\ref{poirep},\ref{covardev1},\ref{covardev2}).

\section{The induced $(1+1)$-Minkowski graded super-Poincar\'e algebra}

One of the consequences of the  graded superspace of the worldline \zzg super-Poincar\'e algebra
${\cal P}$ is that it accommodates a two-dimensional Minkowski \zzg super-Poincar\'e algebra ${\cal P}_{d=2}$ which presents the addition of an extra generator
(the Lorentz boost). The exotic bosonic coordinate is the responsible for the extra space coordinate $y$ given in  (\ref{matrixcoor}). \par
The induced  $5$-generator ${\cal P}_{d=2}$ superalgebra is a new \zzg extension of the two-dimensional Poincar\'e algebra which {\it differs} from the two previous 2D \zzg super-Poincar\'e algebras that have been discussed in the literature. In \cite{tol} a \zzg extension with $11$-graded translations was presented; a different 
\zzg 2D super-Poincar\'e algebra with two $00$-graded translations was introduced in \cite{brusigma}. 
\\
The assignment of the ${\mathbb Z}_2\times{\mathbb Z}_2$-gradings of the ${\cal P}_{d=2}$ generators is, on the other hand, given by
\bea
&&{\textrm{00-graded sector: ~~1 translation,}}\nonumber\\
&& {\textrm{11-graded sector: ~~1 translation and 1 Lorentz boost,}}\nonumber\\
&&{\textrm{10-graded sector: ~~1 parafermion,}}\nonumber\\
&&{\textrm{01-graded sector: ~~1 parafermion.}}
\eea
In ${\cal P}_{d=2}$ the $00$-translation is proportional to ${\widehat H}$, the $11$-translation is proportional to ${\widehat Z}$, while the parafermions are ${\widehat Q}_{10}$ and ${\widehat Q}_{01}$. Besides these four (\ref{superspacehatop}) operators belonging to ${\cal P}$, the extra generator, the $11$-graded Lorentz boost ${\widehat L}$, is introduced as
\bea\label{boost}
&{\widehat L}= M_3\cdot {\widetilde L},\qquad {\textrm{with}} \quad {\widetilde L}= -iy\partial_t -it\partial_y-\frac{i}{2}\theta\partial_\xi-\frac{i}{2}\xi\partial_{\theta}.
\eea
The closure of the \zzg ${\cal P}_{d=2}\supset {\cal P}$ superalgebra extension is guaranteed by the extra 
(anti)commutators involving ${\widehat L}$; they are given by
\bea
&[{\widehat L},{\widehat P}_0]=i{\widehat P}_1, \quad [{\widehat L},{\widehat P}_1]=i{\widehat P}_0,\quad \{{\widehat L},{\widehat Q}_{10}\}= -\frac{1}{2}{\widehat Q_{01}},\quad \{{\widehat L},{\widehat Q}_{01}\}=\frac{1}{2}{\widehat Q}_{10}.&
\eea
In the above formulas we set as ${\widehat P}_0 ={\widehat H}$ and ${\widehat P}_1=\frac{1}{2}{\widehat Z}$ the generators of the
($00$-graded and, respectively, $11$-graded) translations.\par
This new  \zzg two-dimensional super-Poincar\'e algebra ${\cal P}_{d=2}$  is a genuine extension of the $2D$ Poincar\'e algebra which is reproduced by the $00$-graded and $11$-graded sectors.\par
${\cal P}_{d=2}$ is spanned by the hermitian generators
\bea\label{exotic}
&{\widehat P}_0\in {\cal P}_{d=2;00},\qquad {\widehat P}_1,{\widehat L}\in {\cal P}_{d=2;11},\qquad {\widehat Q}_{10}\in {\cal P}_{d=2;10},\qquad {\widehat Q}_{01}\in {\cal P}_{d=2;01}.&
\eea

It is worth pointing out that,
in the construction of the classical \zzg invariant actions, the use of the covariant derivatives (\ref{superspacecovder2}) automatically implies the invariance under the worldline ${\cal P}$ superalgebra. On the other hand, the invariance under the full two-dimensional
${\cal P}_{d=2}$ superalgebra (that is, the extra generator ${\widehat L}$) has to be imposed as an extra constraint to be satisfied by the Lagrangian
${\cal L}$; the requirement is that ${\widehat L}\cdot {\cal L} \approx 0$ up to total derivatives.

\section{Invariant actions in superspace matrix representation}

Graded superfields and invariant actions are nicely derived in terms of the matrix representation of the \zzg superspace. The graded superfield $\Phi(t, \theta_{10},\theta_{01},z)$ given in (\ref{decomposition})
is mapped into a matrix-valued superfield by taking into account:\\
~\\
$~$ {\it i}) the (\ref{matrixcoor}) matrix representation of the graded superspace coordinates and\\
{\it ii}) the ${\mathbb Z}_2\times{\mathbb Z}_2$ grading of the ${\varphi}_{00},{\widetilde\varphi}_{00},\varphi_{11},{\widetilde\varphi}_{11}, \psi_{10},{\widetilde\psi}_{10},\psi_{01},{\widetilde\psi}_{01}$ component fields
entering the (\ref{decomposition}) decomposition.\\
~\\ In the matrix representation one can consistently set
\bea
\varphi_{00} ={\mathbb I}_4\cdot \varphi_0, \quad \quad {\widetilde \varphi}_{00} ={\mathbb I}_4\cdot {\widetilde \varphi}_0,\quad &&\varphi_{11} =M_3\cdot \varphi_3, \quad {\widetilde \varphi}_{11} =M_3\cdot {\widetilde \varphi}_3,\nonumber\\
  \psi_{10} =M_1\cdot \psi_1, \quad~~ {\widetilde \psi}_{10} =M_1\cdot {\widetilde \psi}_1,~&&\psi_{10} =M_2\cdot \psi_2, \quad {\widetilde \psi}_{01} =M_2\cdot {\widetilde \psi}_2
\eea
in terms of ordinary $(t,x=\frac{1}{4}y^2)$-dependent functions; four of them are bosonic ($\varphi_0,{\widetilde\varphi}_0,\varphi_3,{\widetilde\varphi}_3$) and four of them are fermionic ($\psi_1,{\widetilde\psi}_1,\psi_2,{\widetilde\psi}_2$). By further setting
\bea
&{\varphi}:=\varphi_0+\frac{1}{2}y{\widetilde\varphi}_3, \quad g:= -\varphi_3-\frac{1}{2}{\widetilde\varphi}_0,\quad
\psi := -\psi_1+\frac{1}{2}{\widetilde\psi}_2,\quad\chi := -\psi_2-\frac{1}{2}{\widetilde\psi}_1&
\eea
we have that $\Phi(t, \theta_{10},\theta_{01},z)$ is mapped into the $4\times 4$ identity matrix times an ordinary
${\cal N}=2$ bosonic superfield $B(t,y,\theta,\xi)$. We have
\\~
\bea\label{matrixmap}
\Phi(t, \theta_{10},\theta_{01},z)&\mapsto &{\overline \Phi}({\overline t}, {\overline \theta}_{10},{\overline \theta}_{01},{\overline z})={\mathbb I}_4\cdot B(t,y,\theta,\xi), \qquad {\textrm{where}}\nonumber\\
&&~~~~~B(t,y,\theta,\xi) =\varphi (t,y)-i\theta \psi(t,y)-i\xi\chi(t,y)-i\theta\xi g(t,y).
\eea
~\\
After setting the infinitesimal parameters (\ref{gleft})  of the \zzg worldline super-Poincar\'e algebra to be 
\bea
&\varepsilon_{00}=\varepsilon_0\cdot {\mathbb I}_4, \qquad \varepsilon_{10}=\varepsilon_1\cdot M_1,\qquad \varepsilon_{01}=\varepsilon_2\cdot M_2,\qquad \varepsilon_{11}=\varepsilon_3\cdot M_3&
\eea
and the infinitesimal parameter $\varepsilon_{L}$ of the $-i\varepsilon_L{\widehat L}$ Lorentz boost (\ref{boost})  to be
\bea
\varepsilon_{L} &=&\varepsilon_4\cdot M_3,
\eea
the infinitesimal  transformations of the ${\cal P}_{d=2}$ superalgebra (\ref{exotic}) on the $\varphi,\psi,\xi,g$
component fields read as follows:
\bea\label{transformations}
\delta\varphi&=&\varepsilon_{0}{\dot{\varphi}-i\varepsilon}_1\psi-i\varepsilon_2\chi+2\varepsilon_3\varphi' -\varepsilon_4(y{\dot\varphi} +t\varphi'),\nonumber\\
\delta\psi&=&\varepsilon_{0}{\dot{\psi}-\varepsilon}_1{\dot\varphi}-\varepsilon_2(g+\varphi')+2\varepsilon_3\psi' -\varepsilon_4(y{\dot\psi} +t\psi'+\frac{1}{2}\chi),\nonumber\\
\delta\chi&=&\varepsilon_{0}{\dot{\chi}+\varepsilon}_1(g-\varphi')-\varepsilon_2{\dot\varphi}+2\varepsilon_3\chi' -\varepsilon_4(y{\dot\chi} +t\chi'+\frac{1}{2}\psi),\nonumber\\
\delta g&=&\varepsilon_{0}{\dot{g}+i\varepsilon}_1({\dot\chi}-\psi')-i\varepsilon_2({\dot\psi}-\chi')+2\varepsilon_3\varphi' -\varepsilon_4(y{\dot g} +tg').
\eea
The infinitesimal parameters $\varepsilon_0,\varepsilon_3,\varepsilon_4$  are real while $\varepsilon_1,\varepsilon_2$ 
are Grassmann numbers; in the above formulas $f'\equiv \partial_y f$.\par
~\par
The graded superspace integration can be introduced in the matrix representation through the position
\bea
\int dt dz d\theta_{01}d\theta_{10}  &=&\frac{1}{8} Tr \int dtdy d\xi d\theta \cdot M_3M_2M_1= -\frac{i}{8} Tr\int dtdy d\xi d\theta \cdot {\mathbb I}_4.
\eea
This integration is equivalent to the prescription (\ref{inv12221}) that was discussed in Section {\bf 7}. It follows that, by construction, the actions constructed with potential terms and the covariant derivatives 
${\widehat D}_{10}, {\widehat D}_{01}$ are invariant under the \zzg worldline super-Poincar\'e algebra.  \par
~\par
An invariant action ${\cal S}$ is given by
\bea\label{invaction1}
{\cal S} &=& \int dt dz d\theta_{01} d\theta_{10} \cdot {\cal L}, \qquad {\textrm{where}}\nonumber\\
{\cal L} &=& \gamma_{11} {\widehat D}_{10}\Phi {\widehat D}_{01}\Phi -V(\Phi).
\eea
The constant $\gamma_{11}$ is assumed to be $11$-graded in order to have a $00$-graded action  ${\cal S}$. 
By setting
$\gamma_{11}= \gamma M_3$, with $\gamma\in{\mathbb R}$, the matrix representation of the action ${\cal S}$ is given by
\bea
{\cal S}&=&-\frac{i}{2}\int dt dy d\xi d\theta \big(i\gamma {\widetilde D}_1B\cdot {\widetilde D}_2B-V(B)\big).
\eea
In terms of the component fields, after integrating in $\xi,\theta$ and taking the $\gamma=-1$ value, we get
\bea\label{invaction2}
{\cal S}&=& \int dt dy \big({\cal L}_{kin} -{\cal L}_{pot}\big), \qquad{\textrm{where}}\nonumber\\
{\cal L}_{kin} &=& \frac{1}{2}\big({\dot\varphi}^2 -(\varphi')^2+g^2 +i\psi{\dot\psi}+i\chi{\dot\chi} -i\psi\chi'-i\chi\psi'\big),\nonumber\\
{\cal L}_{pot} &=&- \frac{1}{2}\big(gV_\varphi+i\psi\chi V_{\varphi\varphi}\big), \qquad {\textrm{for}}~~ V_\varphi\equiv \frac{\partial V}{\partial \varphi}.
\eea
The Euler-Lagrange equations are
\bea\label{eulerlag}
{\ddot\varphi}-\varphi''-\frac{1}{2}gV_{\varphi\varphi}-\frac{i}{2}\psi\chi V_{\varphi\varphi\varphi}=0,&\qquad\qquad &  g+\frac{1}{2}V_{\varphi}=0,\nonumber\\
{\dot\psi}-\chi'+\frac{1}{2}\chi V_{\varphi\varphi}=0,\qquad\qquad&\qquad\qquad&
{\dot\chi}-\psi'-\frac{1}{2}\psi V_{\varphi\varphi}=0.\qquad
\eea
It follows that $g(t,y)$ is an auxiliary field whose equation of motion can be algebraically solved.\par
~\par
As remarked at the end of Section {\bf 9}, the action (\ref{invaction1}) is invariant by construction under the worldline \zzg super-Poincar\'e algebra ${\cal P}$. In order for the system to describe a two-dimensional relativistic theory, the action (\ref{invaction1}) should also be invariant under the Lorentz boost ${\widehat L}$. It is easily proved, with lengthy but straightforward computations, that this is indeed the case for any choice of the  potential $V(\Phi)$. Therefore, the action  (\ref{invaction1}) is invariant under the full two-dimensional \zzg ${\cal P}_{d=2}$ super-Poincar\'e algebra (\ref{exotic}). The demonstration requires to check that applying the Lorentz boost ${\widetilde L} $ from (\ref{boost}) to the Lagrangian $ {\cal L}={\cal L}_{kin}-{\cal L}_{pot}$ of (\ref{invaction2}), a total derivative is produced. We just present here the sketch of the demonstration.\par
The computation 
\bea\label{upto}
{\widetilde L}\cdot{\cal L}&\approx & 0\qquad ({\textrm{up to total derivatives}}) 
\eea
is split into different terms which should separately verify (\ref{upto}): the purely bosonic part, the part which contains bilinear in $\psi$ contributions, the part with the bilinear in $\chi$ contributions and, finally, the part with ``mixed" $\psi,\chi$ fermionic contributions. As an example, the bilinear in $\psi$ term from ${\widetilde L}\cdot{\cal L}_{kin}$ is proportional to $2t\psi'{\dot\psi}-\psi\psi'$; this term is equivalent to
$t\psi{\dot\psi}'-t{\dot\psi}\psi'$ after integrating by part and, furthermore, to $\partial_y(t\psi{\dot\psi})$. A similar analysis, conducted on the other terms, prove the validity of (\ref{upto}) for any choice of the potential.\par
By applying standard methods, see e.g. \cite{husk}, we can present the conserved currents and Noether charges associated with the $5$ invariant  symmetry operators ${\widehat H}, {\widehat Z}, {\widehat L}, {\widehat Q}_{10}, {\widehat Q}_{01}$ of the (\ref{invaction1}) action ${\cal S}$. Under the (\ref{matrixmap}) mapping and the (\ref{superspacehatop},\ref{boost}) positions, the conserved currents
and charges are reconstructed from the corresponding ones obtained from the invariant operators ${\widetilde H}, {\widetilde Z}, {\widetilde L}, {\widetilde Q}_{1}, {\widetilde Q}_2$ of (\ref{invaction2}) acting on the $B(t,y,\theta,\xi)$ superfield. \par
The five conserved currents 
$J_\ast^\mu$ (for $\ast\equiv {\widetilde H}, {\widetilde Z}, {\widetilde L}, {\widetilde Q}_{1}, {\widetilde Q}_{2}$ and $\mu\equiv t,y$) are defined  as
\bea
J_\ast^\mu&=&\Lambda^\mu -\delta\Phi_A\frac{\partial {\cal L}}{\partial(\partial_\mu\Phi_A)}, \qquad {\textrm{where}}\quad \delta {\cal L} = \partial_\mu\Lambda^\mu\quad{\textrm{and}}\quad \Phi_A\equiv \varphi,\psi,\chi,g.
\eea
By taking into account the equations of motion, the currents satisfy the conserved equations
\bea
\partial_tJ_\ast^t+\partial_yJ_\ast^y&=&0.
\eea 
The computation of the conserved currents gives
\bea
J_{\widetilde H}^t &=&-\frac{1}{2}\big({\dot{\varphi}}^2+(\varphi')^2+i\psi\chi'+i\chi\psi'+\frac{1}{4}V_{\varphi}^2-i\psi\chi V_{\varphi\varphi}\big) ,\nonumber\\ J_{\widetilde H}^y &=& 
{\dot{\varphi}}\varphi'+\frac{i}{2}(\chi {\dot\psi}+\psi{\dot\chi});\nonumber\\
J_{\widetilde Z}^t&=&-2{\dot\varphi}\varphi'-i\psi\psi'-i\chi\chi',\nonumber\\ 
J_{\widetilde Z}^y &=& {\dot\varphi}^2+(\varphi')^2+i\psi{\dot\psi}+i\chi{\dot \chi}-\frac{1}{4}V_\varphi^2+i\psi\chi V_{\varphi\varphi};\nonumber\\
J_{\widetilde L}^t&=&\frac{y}{2}\big( {\dot\varphi}^2+(\varphi')^2+i(\psi\chi'+\chi\psi')+\frac{1}{4}V_\varphi^2-i\psi\chi V_{\varphi\varphi}\big) +t\big({\dot\varphi}\varphi'+\frac{i}{2}(\psi\psi'+\chi\chi')\big),\nonumber\\
J_{\widetilde L}^y &=&\frac{t}{2}\big( -{\dot\varphi}^2-(\varphi')^2-i(\psi{\dot\psi}+\chi{\dot\chi})+\frac{1}{4}V_\varphi^2-i\psi\chi V_{\varphi\varphi}\big)  +y\big(-{\dot\varphi}\varphi'-\frac{i}{2}(\psi{\dot\chi}+\chi{\dot\psi})\big);\nonumber\\
J_{{\widetilde Q}_{1}}^t&=&-{\dot\varphi}\psi-\varphi'\chi-\frac{1}{2}\chi V_\varphi ,\nonumber\\
J_{{\widetilde Q}_{1}}^y&=&{\dot\varphi}\chi+\varphi'\psi+\frac{1}{2}\psi V_\varphi ;\nonumber\\
J_{{\widetilde Q_{2}}}^t&=&-{\dot\varphi}\chi-\varphi'\psi+\frac{1}{2}\psi V_\varphi  ,\nonumber\\
J_{{\widetilde Q_{2}}}^y&=&{\dot\varphi}\psi+\varphi'\chi-\frac{1}{2}\chi V_\varphi  .
\eea
As usual, the Noether charges are obtained from the integration $\int dy J_\ast^t$.\par
~\par
{\bf Remarks.} The following remarks summarize the results of this Section: \\
~\par
1 - the  superspace of the
worldline \zzg super-Poincar\'e algebra induces a two-dimensional relativistic model which is invariant under the 
two-dimensional \zzg Poincar\'e algebra ${\cal P}_{d=2}$. The action ${\cal S}$ in (\ref{invaction1}) 
is invariant under this larger algebra; \par
~\par
2 - under the (\ref{matrixmap}) mapping the \zzg model is mapped into an ordinary  two-dimensional relativistic (${\cal N}=1$) theory defined for the scalar bosonic superfield $B(t,y,\theta,\xi)$. The \zzg invariant operators and conserved charges are reconstructed from the corresponding supersymmetric invariant operators and charges.

\section{The \zzg closed string}

The compactification of the $y$ coordinate on the ${\bf S}^1$ circle parametrized by $y\in[0,2\pi[$, coupled with the (\ref{compactintegral}) integration,
produces a \zzg closed string. We present its construction. \par
We note at first that the ${\mathbb Z}_2\times{\mathbb Z}_2$ grading allows both periodic ($P$) and/or antiperiodic ($A$) boundary conditions. By assuming the ordinary $00$-graded bosons to be periodic, the consistency of the  $mod~2$ grading
addition given in (\ref{mod2sum}) requires the following (anti)periodicity of the remaining graded sectors. One of the three alternatives
are in principle admissible  in association with the corresponding $(00/10/01/11)$ graded sectors:
\bea\label{ppandaa}
&i: (P/P/P/P), \qquad ii: (P/A/A/P),\qquad iii: (P/P/A/A)\equiv (P/A/P/A).&
\eea

Even if the three above alternatives are admissible by the ${\mathbb Z}_2\times{\mathbb Z}_2$ grading, it does not mean
that they are necessarily implemented in a specific model.  As an example, inspecting the transformations parametrized by $\varepsilon_1,\varepsilon_2$, the matching of left and right hand sides periodicities in (\ref{transformations}) requires the fermions $\psi,\chi$ to be periodic. Similarly, the auxiliary field $g$ needs to be periodic. It follows that the implementation of the \zzg worldline super-Poincar\'e algebra for a closed string 
requires all component fields to be periodic. \par
This analysis does not rule out  the possibility of antiperiodic boundary conditions for the $y$-compactified closed string recovered from the (\ref{invaction2}) Lagrangian. By inspecting the (\ref{eulerlag}) equations of motion one can
for instance check that, for a vanishing potential $V(\varphi)=0$, the $g=0$ equation is compatible with  an antiperiodic boundary condition for the auxiliary field $g$. \par
We limit here to discuss the  closed string with $(P/P/P/P)$ boundary conditions since this is the case which is directly derived from the \zzg superspace formalism (comments about the two other possible options in (\ref{ppandaa}) are given later).\par
Let us assume, for the (\ref{invaction1}) action which leads to (\ref{invaction2}), the periodicity conditions for the 
component fields:
\bea
&\varphi(y+2\pi)=\varphi(y), \quad \psi(y+2\pi)=\psi(y),\quad \chi(y+2\pi)=\chi(y),\quad g(y+2\pi)=g(y).&
\eea
The given component field $f(t,y)$ (where $f$ denotes any of the above $\phi,\psi,\chi, g$ fields) is real and mode-expanded
according to
\bea
f(t,y)&=&\sum_{n=-\infty}^{+\infty} f_n(t) e^{iny}, \qquad {f_n^\ast=f_{-n}}.
\eea 
The $y$-integration in (\ref{invaction2}) is then defined as
\bea
\int dy &\equiv& \frac{1}{2\pi}\int_0^{2\pi} dy.
\eea
After performing the integration in $y$ the action (\ref{invaction2}) reads as
\bea\label{closedstring}
{\cal S} &=& \int dt \big( {\overline{\cal L}}_{kin}-{\overline{\cal L}}_{pot}\big).
\eea
The kinetic term is
\bea
{\overline{\cal L}}_{kin} &=& \frac{1}{2}\sum_{n=-\infty}^{+\infty}\big(|{\dot\varphi}_n|^2-n^2|\varphi_n|^2+|g_n|^2+i\psi_{n}{\dot\psi}^\ast_n+i\chi_{n}{\dot\chi}_n^\ast +2n{\psi}^\ast_n\chi_n\big).
\eea
If we take the potential term to be obtained from a quadratic $V(\Phi)\propto \Phi^2$ in (\ref{invaction1}) we get
\bea
{\overline{\cal L}}_{pot} &=&\sum_{n=-\infty}^{+\infty}k\big(g_n\varphi_n^\ast+i\psi_n\chi_n^\ast\big)\qquad{\textrm{for}\quad k\in {\mathbb R}.}
\eea
The Euler-Lagrange equations of motion are
\bea
{\ddot \varphi}_n+n^2\varphi_n+kg_n=0,&\quad& g_n-k\varphi_n=0,\nonumber\\
i{\dot\psi}_n+n\chi_n-ik\chi_n=0,&\quad&i{\dot\chi}_n+n\psi_n+ik\psi_n=0.
\eea
We get in particular, after solving the equations of motion for the auxiliary fields $g_n(t)$:
\bea
&{\ddot \varphi}_n= -(n^2+k^2)\varphi_n, \qquad{\ddot \psi}_n= -(n^2+k^2)\psi_n, \qquad{\ddot \chi}_n= -(n^2+k^2)\chi_n.&
\eea
The following three scenarios apply:\par
{\it i}) the free closed string (for $V(\Phi)=0\Rightarrow k=0$) corresponds to an infinite set of graded harmonic oscillators for $n\neq 0$ plus the zero-modes recovered from $\varphi_0,\psi_0,\chi_0$;\par
{\it ii}) for the quadratic potential ($k\neq 0$) the energy of the harmonic oscillators is shifted. The fields $\varphi_0,\psi_0,\chi_0$ no longer describe zero-modes, but harmonic oscillators with energy $|k|$;
\par
{\it iii}) for an arbitrary potential $V(\Phi)$ the \zzg invariant action produces an infinite set of interacting particles.\par
In scenarios {\it i} and {\it ii} one can consistently describe the dynamics of the $n$-th mode component fields alone.
This situation corresponds to restrict the two-dimensional $8\times 8$ matrix $D$-module representation (\ref{reldmod}) to the $4\times 4$ worldline $D$-module representation (\ref{121beta}). This is no longer the case in scenario {\it iii} for an arbitrary $V(\Phi)$. The models with interacting modes cannot be restricted to the irreducible time-dependent $4$-component fields, so that in the interacting case the full $D$-module representation (\ref{reldmod}) is required. This leads to a new class of \zzg invariant models not previously considered in the literature.\par
~\par
We conclude this Section by pointing out some open questions, left for future investigations, concerning the (anti)periodic boundary conditions for the
\zzg closed string. In \cite{zhe} a para-Grassmann string (different from our model) was presented for both Ramond
(periodic, ``R") and Neveu-Schwarz (antiperiodic, ``NS") boundary conditions, the two versions being related by a Klein transformation. This suggests that, for the free closed string model (\ref{closedstring}) with ${\cal L}_{pot}=0$, the analysis of the invariance  should be conducted for the separate left-right mover sectors given by $x_\pm =t\pm y$ in all four cases: $R$-$R$, $R$-$NS$, $NS$-$R$, $NS$-$NS$. This analysis, which will be presented in a forthcoming paper, should also lead to \zzg extensions of the Virasoro algebra.

\section{Conclusions}
\par
Here we summarize and comment some of the results of the paper.\par
It has been shown that the superspace of the worldline \zzg super-Poincar\'e algebra induces $D$-module representations
acting on $8$ two-dimensional component fields. Rather unexpectedly the derived classical actions, see e.g. formula (\ref{invaction1}), are invariant under a two-dimensional relativistic \zzg super-Poincar\'e algebra. This two-dimensional superalgebra and the associated  relativistic models have not been previously considered in the literature. Furthermore, the compactification of the second (space) coordinate on a circle produces  a \zzg closed string theory
with periodic boundary conditions. A potential term, producing interacting modes, can be consistently added.\par
The irreducibility conditions of the \zzg supermultiplets have been analyzed. We proved  that the previously
known $4$-component worldline supermultiplets admit  (see the presentations in Section {\bf 6}) a $\beta$-deformation, the original worldline supermultiplets being recovered at $\beta=0$; this is a special supersymmetric point (see formula (\ref{susycritical}), where $\alpha= \frac{1}{2}+\beta$).\par
A useful presentation of the \zzg calculus, in terms of matrices encoding the ${\mathbb Z}_2\times{\mathbb Z}_2$-grading coupled with the Berezinian calculus,  has also been introduced in Section {\bf 8}.\par
These results imply that a larger class of \zzg invariant models (both two-dimensional and one-dimensional) than the ones so far considered, becomes available. These models fall into the realm of \zzg parastatistics with, see \cite{{top1},{top2}}, detectable
consequences for the presence of the paraparticles.\par
It prompts to further investigations of these theories in both classical and quantum settings. For closed string models,
see the remarks at the end of Section {\bf 11}, this means investigating \zzg strings with Ramond and Neveu-Schwarz
boundary conditions and \zzg extensions of the Virasoro algebra. The construction of $\beta\neq 0$ deformed worldline models is left for a forthcoming paper. Unlike the $\beta=0$ theories of references \cite{{brdu},{akt1},{akt2}}, the energy spectrum of these deformed theories is not related by a supersymmetry transformation.

~

  \renewcommand{\theequation}{A.\arabic{equation}}
  \setcounter{equation}{0} 
\par
~\\
{\bf{\Large{Appendix A: irreducible representations and $\beta$-deformation of the supersymmetric spectrum}}}
~\par
~\par
The irreducible representations of the \zzg one-dimensional super-Poincar\'e algebra (\ref{z2z2poin}) on a \zzg space are $4$-dimensional. On the other hand, in physical applications, in order to derive the energy spectrum of the invariant models is also important to consider representations on ${\mathbb Z}_2$-graded spaces. The irreducible ones are $2$-dimensional. The  analysis goes as follows.
\par
The four hermitian generators of (\ref{z2z2poin}) are $H, Z,Q_{10}, Q_{01}$.  We set, for simplicity, 
\bea
Q_1 := Q_{10}, && Q_2:= Q_{01}.
\eea
In the enveloping algebra we can introduce the operator
$W$, hermitian under the $\ast$ conjugation defined in (\ref{star}),  which is expressed by the anticommutator
\bea
W &:=& \{Q_1,Q_2\}, \qquad\qquad(W^\ast=W).
\eea
A simple computation  shows that
\bea
\relax &[W,Q_1]=[W,Q_2]=0.&
\eea
It is convenient to set
\bea
Q_\pm :={\footnotesize{ \frac{1}{\sqrt{2}}}}(Q_1\pm Q_2)\quad &{\textrm{and}}&\quad  H_\pm := H\pm \frac{1}{2}W.
\eea
The ${\mathbb Z}_2$-graded $4$-generator superalgebra with $Q_\pm$ odd and $H_\pm$ even elements is recovered:
\bea
&\{Q_\pm, Q_\pm\}=2H_\pm,\qquad  \{Q_+,Q_-\}=0,\qquad  [H_\pm, G]=0 \quad{\textrm{for any}} \quad G= H_\pm, Q_\pm.&
\eea
The above superalgebra gives two independent copies (for $Q_+,H_+$ and $Q_-,H_-$, respectively)  of the one-dimensional ${\cal N}=1$ supersymmetry.
For $W=0$ one gets the $3$-generator one-dimensional ${\cal N}=2$ supersymmetry defined by $Q_\pm$ and  
$H_{susy}\equiv H_+=H_-$:
\bea\label{n2susy}
\{Q_\pm, Q_\pm\}=2H_{susy},&& [H_{susy}, Q_\pm]=0.
\eea
\par
Following the presentation of Section {\bf 5} we can parametrize the $Z^2$ Casimir operator of (\ref{z2z2poin}) as $Z^2=\lambda^2$ by assuming $\lambda >0$.  The energy eigenvalue, corresponding to the Casimir operator $H$, can be parametrized, see (\ref{ealphalambda}), as $\alpha\lambda$ where $\alpha$ is a nondimensional real parameter. Since $H$ is the square of hermitian operators, $\alpha$ is a non-negative real number. The following analysis proves that 
$\alpha$ is restricted to satisfy
\bea\label{alphaconstr}
\alpha &\geq & \frac{1}{2}.
\eea
The relation
\bea
&W^2 = -Z^2+4H^2 \quad {\textrm{implies that}}\quad 
W^2 = (4\alpha^2-1)\lambda^2.&
\eea
By taking into account the hermiticity of $W$, the (\ref{alphaconstr}) constraint on $\alpha$ follows.
\\
The eigenvalues of $W$ are $\lambda\sqrt{4\alpha^2-1}$; therefore the degenerate $E_\pm$ eigenvalues of $H_\pm$ are
\bea
E_\pm &=&\lambda\big(\alpha \pm\frac{1}{2}\sqrt{4\alpha^2-1}\big).
\eea
Since the $W=0$ condition is obtained for $\alpha=\frac{1}{2}$, it is convenient 
 to express $\alpha$ in terms of the shifted parameter $\beta$:
\bea\label{alphabeta}
\alpha=\beta+\frac{1}{2}, && {\textrm{where}} \quad \beta\geq 0.
\eea

The boundary point $\beta=0$ corresponds to the critical value which produces the enhanced ${\cal N}=2$ supersymmetry (\ref{n2susy}).\par
In terms of $\beta$ the $E_\pm$ eigenvalues are
\bea\label{epmbeta}
E_\pm &=&\lambda\big(\frac{1}{2}+\beta \pm\sqrt{\beta^2+\beta}\big).
\eea
The $E_\pm(\beta)$ functions are strictly monotonic (respectively, crescent/decrescent) in the $\beta\geq 0$ domain, with
\bea
E_\pm (\beta=0) =\frac{1}{2},\quad  &{\textrm{while}}& \lim_{~~~\beta\rightarrow +\infty}E_+(\beta) =+\infty, \qquad
\lim_{~~\beta\rightarrow +\infty}E_-(\beta) =0.\quad
\eea
A $4$-dimensional representation of the ${\mathbb Z}_2$-graded superalgebra (\ref{z2z2poin}), labeled by the pair of eigenvalues $(E_+, E_-)$  given by
\bea
H_\pm |vac\rangle &=& E_\pm |vac\rangle,
\eea
is spanned by the $4$ vectors 
\bea
&|vac\rangle, \qquad Q_+|vac\rangle,\qquad Q_-|vac\rangle, \qquad Q_+Q_-|vac\rangle.&
 \eea
This representation is reducible. By taking into account the ${\mathbb Z}_2$-graded structure, a $2$-dimensional  irreducible representation is recovered by constraining
\bea
Q_+Q_-|vac\rangle = b|vac\rangle, && Q_-|vac\rangle = f Q_+|vac\rangle.
\eea
The constants $b,f$ are determined by the consistency conditions
$H_-|vac\rangle=Q_-Q_-|vac\rangle= fQ_-Q_+|vac\rangle$ and $H_+|vac\rangle=Q_+Q_+|vac\rangle= \frac{1}{f}Q_+Q_-|vac\rangle$. One gets
\bea
b =i\sqrt{E_+E_-}, &\qquad& f=i\sqrt{\frac{E_-}{E_+}}.
\eea
The action of $Q_\pm$ on the $2$-component vector $v^T = (|vac\rangle, Q_+|vac\rangle)$ gives the $2\times 2$ matrix representation 
\bea
&{\overline H}_\pm = E_\pm\cdot {\mathbb I}_2,\qquad
{\overline Q}_+ =\left(\begin{array}{cc} 0&E_+\\1&0\end{array}\right), \qquad
{\overline Q}_- =\left(\begin{array}{cc} 0&-i\sqrt{E_+E_-}\\i\sqrt{\frac{E_-}{E_+}}&0\end{array}\right).&
\eea  
The above operators produce, by inserting the (\ref{epmbeta}) relations for $E_\pm(\beta)$, a class of $\beta$-dependent $2$-dimensional representations.
~\par
~\par
\par {\Large{\bf Acknowledgments}}
{}~\par{}~\par

Z. K. and F. T. are grateful to the Osaka Metropolitan University, where this work was initiated, for hospitality.
 F. T.  has been supported by CNPq (PQ grant 308846/2021-4).


\begin{thebibliography}{99}
\bibitem{bruz2n} A. J. Bruce, {\it{On a $\mathbb{Z}_2^n$-graded version of supersymmetry}}, Symmetry {\bf 11}, 116 (2019); arXiv:1812.02943 [hep-th]. 
\bibitem{tol} V. N. Tolstoy, {\it Super-de Sitter and alternative super-Poincar\'e symmetries}, in ``Lie Theory and Its Application in Physics" (Ed. V. Dobrev), Springer Proceedings in Mathematics, {\bf v. 111}, 357 (2014); 
arXiv:1610.01566[hep-th].
\bibitem{riwy1} V. Rittenberg and D. Wyler, {\it Generalized Superalgebras}, {Nucl. Phys.} {\bf B 139}, 189 (1978).
\bibitem{riwy2} V. Rittenberg and D. Wyler, {\it Sequences of $Z_2\otimes Z_2$ graded Lie algebras and superalgebras}, {J. Math. Phys.} {\bf 19}, 2193 (1978).
\bibitem{sch1} M. Scheunert, {\it Generalized Lie algebras}, {J. Math. Phys.} {\bf 20}, 712 (1979).
\bibitem{kuto} Z. Kuznetsova and F. Toppan, {\it Classification of minimal \zzg Lie (super)algebras and some applications}, J. Math. Phys. {\bf 62}, 063512 (2021); arXiv:2103.04385[math-ph].
\bibitem{kac}  V. G. Kac, {\it Lie Superalgebras}, Adv. in Math. {\bf 26}, 8 (1977).
\bibitem{luri} J. Lukierski and V. Rittenberg, {\it Color-De Sitter and Color-Conformal Superalgebras}, {{Phys. Rev.}} {\bf D 18},  385 (1978).
\bibitem{vas} M. A. Vasiliev, {\it de Sitter supergravity with positive cosmological constant and generalized Lie superalgebras}, {{Class. Quantum Grav.}} {\bf 2},  645 (1985).
\bibitem{jyw} P. D. Jarvis, M. Yang and B. G. Wybourne, {\it Generalized quasispin for supergroups}, {{J. Math. Phys.}} {\bf 28},  1192 (1987).
\bibitem{aktt1} N. Aizawa, Z. Kuznetsova, H. Tanaka and F. Toppan, {\it $ \mathbb{Z}_2 \times \mathbb{Z}_2$-graded Lie symmetries of the L\'evy-Leblond equations}, {Prog. Theor. Exp. Phys.} {\bf{2016}}, 123A01 (2016); arXiv:1609.08224[math-ph].
\bibitem{aktt2} N. Aizawa, Z. Kuznetsova, H. Tanaka and F. Toppan, {\it Generalized supersymmetry and L\'evy-Leblond equation}, in S. Duarte {et al} (eds), {Physical and Mathematical Aspects of Symmetries}, Springer, Cham, p. 79 (2017); arXiv:1609.08760[math-ph].
\bibitem{akt1} N. Aizawa, Z. Kuznetsova and F. Toppan, {\it ${\mathbb Z}_2\times{\mathbb Z}_2$-graded mechanics: the classical theory}, Eur. J. Phys. {\bf C 80}, 668 (2020); arXiv:2003.06470[hep-th].
\bibitem{akt2} N. Aizawa, Z. Kuznetsova and F. Toppan, {\it ${\mathbb Z}_2\times {\mathbb Z}_2$-graded mechanics: the quantization}, Nucl. Phys. {\bf B 967}, 115426 (2021); arXiv:2005.10759[hep-th].
\bibitem{brusigma} A. J. Bruce, {\it ${\mathbb Z}_2\times{\mathbb Z}_2$-graded supersymmetry: 2-d sigma models}, {{ J. Phys. A: Math. Theor. {\bf 53}, 455201 (2020)}}; arXiv:2006.08169[math-ph].
\bibitem{bruSG} A. J. Bruce, {\it Is the $\mathbb{Z}_2 \times \mathbb{Z}_2$-graded sine-Gordon equation integrable?}, Nucl. Phys. B {\bf 971}, 115514 (2021); arXiv:2106.06372 [math-ph].
\bibitem{brdu} A. J. Bruce and S. Duplij, {\it Double-graded supersymmetric quantum mechanics}, {J. Math. Phys.} {\bf 61}, 063503 (2020); arXiv:1904.06975 [math-ph].	
\bibitem{AAD} N. Aizawa, K. Amakawa, S. Doi, {\it $\mathcal{N}$-Extension of double-graded supersymmetric and superconformal quantum mechanics}, J. Phys. A: Math. Theor. \textbf{53}, 065205 (2020); arXiv:1905.06548 [math-ph].
\bibitem{DoiAi1} S. Doi and N. Aizawa, {\it $\mathbb{Z}_2^3$-Graded extensions of Lie superalgebras and superconformal quantum mechanics}, SIGMA {\bf 17}, 071 (2021); arXiv:2103.10638 [math-ph]. 
\bibitem{AAd2} N. Aizawa, K. Amakawa, S. Doi, {\it $\mathbb{Z}_2^n$-Graded extensions of supersymmetric quantum mechanics via Clifford algebras}, J. Math. Phys. {\bf 61}, 052105 (2020); arXiv:1912.11195 [math-ph]. 
\bibitem{top1} F. Toppan, {\it \zzg parastatistics in multiparticle quantum Hamiltonians}, J. Phys. A: Math. Theor. {\bf 54}, 115203 (2021); arXiv:2008.11554[hep-th].
\bibitem{top2} F. Toppan, {\it Inequivalent quantizations from gradings and \zzg parabosons}, J. Phys. A: Math. Theor.  {\bf 54}, 355202 (2021); arXiv:2104.09692[hep-th].
\bibitem{YJ} W. M. Yang and S. C. Jing, {\it A new kind of graded Lie algebra and parastatistical supersymmetry}, Sci. in China (Series A) \textbf{44}, 9 (2001); arXiv:math-ph/0212004. 
\bibitem{tol1} V. N. Tolstoy, {\it Once more on parastatistics}, {Phys. Part. Nucl. Lett.} {\bf{11}},  933 (2014); arXiv:1610.01628[math-ph].
\bibitem{stvj1} N. I. Stoilova and J. Van der Jeugt, {\it The ${\mathbb Z}_2\times{\mathbb Z}_2$-graded Lie superalgebra $pso(2m+1|2n)$ and new parastatistics representations},  J. Phys. {A}: Math. Theor. {\bf 51}, 135201 (2018); arXiv:1711.02136[math-ph].
\bibitem{AmaAi}  K. Amakawa and N. Aizawa, {\it A classification of lowest weight irreducible modules over $\mathbb{Z}_2^2$-graded extension of $osp(1|2)$}, J. Math. Phys.  \textbf{62}, 043502 (2021); arXiv:2011.03714 [math-ph].
\bibitem{que} C. Quesne, {\it Minimal bosonization of double-graded supersymmetric quantum mechanics}, Mod. Phys. Lett. {\bf A 36}, 2150238 (2021); arXiv:2108.06243 [math-ph]. 
\bibitem{NeliJoris} N. I. Stoilova and J. Van der Jeugt, \textit{The ${\mathbb{Z}}_{2}\times {\mathbb{Z}}_{2}$-graded Lie superalgebras $\mathfrak{p}\mathfrak{s}\mathfrak{o}(2n+1\vert 2n)$ and $\mathfrak{p}\mathfrak{s}\mathfrak{o}(\infty \vert \infty )$, and parastatistics Fock spaces,}  J. Phys. A: Math. Theor. \textbf{55}, 045201 (2022); arXiv:2112.12811 [math-ph]. 
\bibitem{LuTan} R. Lu and Y. J. Tan, {\it Construction of color Lie algebras from homomorphisms of modules of Lie algebras}, J. Algebra {\bf 620}, 1 (2023). 
\bibitem{pon} N. Poncin, {\it Towards integration on colored supermanifolds}, Banach Center Publication {\bf 110}, 201 (2016).
\bibitem{PonSch} N. Poncin, S. Schouten, {\it The geometry of supersymmetry/A concise introduction}; arXiv:2207.12974 [math-ph].  
\bibitem{aido1} S. Doi and N. Aizawa, {\it Comments on ${\mathbb Z}_2^2$-graded supersymmetry in superfield formalism}, Nucl. Phys. {\bf B 974}, 115641 (2022); arXiv:2109.14227[math-ph].
\bibitem{aido2} N. Aizawa and S. Doi, {\it Irreducible representations of ${\mathbb Z}_2^2$-graded ${\cal N}=2$ supersymmetry algebra and ${\mathbb Z}_2^2$-graded supermechanics},  J. Math. Phys. {\bf 63}, 091704 (2022);  arXiv:2205.07263[math-ph].
\bibitem{ber}  F. A. Berezin, {\it The Method of Second Quantization}, Academic Press, (1966).
\bibitem{ckp} T. Covolo, S. Kwok and N. Poncin, {\it{Differential calculus on $\mathbb{Z}_2^n$-supermanifolds}}, arXiv:1608.00949 [math.DG]. 
\bibitem{pato} A. Pashnev and F. Toppan, {\it On the classification of N-extended supersymmetric quantum mechanical systems}, {J. Math. Phys.} {\bf 42}, 5257 (2001); arXiv:hep-th/0010135.
\bibitem{krt} Z. Kuznetsova, M. Rojas and F. Toppan, {\it Classification of irreps and invariants of the $N$-extended supersymmetric quantum mechanics}, JHEP  {\bf{0603}}, 098 (2006); arXiv:hep-th/0511274.
\bibitem{bede} J. Beckers and N. Debergh, {\it On colour superalgebras in parasupersymmetric quantum mechanics}, J. Phys. A: Math. Gen. {\bf 24}, L597 (1991).
\bibitem{cht} I. E. Cunha, N. L. Holanda and F. Toppan, {\it From worldline to quantum superconformal mechanics with and without oscillatorial terms: $D(2,1;\alpha)$ and $sl(2|1)$ models}, {Phys. Rev.} {\bf{D 96}}, 065014 (2017); arXiv:1610.07205[hep-th].
\bibitem{husk} T. Hurth and K. Skenderis, {\it Quantum Noether Method}, Nucl. Phys. {\bf B 541}, 566 (1999); arXiv:hep-th/9803030.
\bibitem{zhe} A. A. Zheltukhin, {\it Para-grassmann extension of the Neveu-Schwarz-Ramond algebra}, Teor. Mat. Fiz {\bf 71}, 218 (1987); Theor. Math. Phys. {\bf 71}, 491 (1987) (English version).
\end{thebibliography}
\end{document}